\documentclass[draftclsnofoot, onecolumn, 12pt]{IEEEtran}
\usepackage{color}
\usepackage{graphicx}
\usepackage{amsmath}
\usepackage{amsfonts}
\usepackage{amssymb}
\usepackage{booktabs}
\usepackage{cases}
\usepackage{tabularx}
\usepackage{boxedminipage}
\usepackage{url}
\usepackage{xcolor}
\usepackage{subcaption}

\begin{document}

\title{\huge Capacity of the Two-Hop  Relay Channel with Wireless Energy Transfer  from   Relay to  Source and Energy Transmission Cost}
\author{ Nikola Zlatanov, Derrick Wing Kwan Ng, and Robert Schober
\thanks{This work has been presented  in part at  IEEE ICC 2016 \cite{conf_vers}.}
\thanks{N. Zlatanov is  with the Department of Electrical and Computer Systems Engineering, Monash University, Melbourne, VIC 3800, Australia (e-mail: nikola.zlatanov@monash.edu).}
\thanks{ D. W. K. Ng  is with the School of Electrical Engineering and Telecommunications, University of New South Wales, Sydney, N.S.W. 2052, Australia (e-mail: w.k.ng@unsw.edu.au).}
\thanks{ R. Schober is with the Friedrich-Alexander University of Erlangen-N\"urnberg, Institute for Digital Communications, D-91058 Erlangen, Germany (e-mail:  robert.schober@fau.de).
}
%\thanks{Digital Object Identif\-ier 10}
\vspace*{-15.5mm}
 }

\maketitle

\begin{abstract}

In this paper, we investigate a communication system comprised of an energy harvesting (EH) source  which harvests radio frequency (RF) energy   from an out-of-band full-duplex relay node and exploits this energy to transmit data to a destination node via the relay node.  
 We assume two scenarios for the battery of the EH source. In the first scenario,   we assume that the EH source is not  equipped with a battery and thereby cannot store energy. As a result, the RF energy  harvested during one symbol interval can only be used  in the following symbol interval. In the second scenario, we assume that the EH source is equipped with a battery having unlimited storage capacity in which it can store the harvested RF energy. As a result, the RF energy  harvested during one symbol interval can  be used  in any of the following symbol intervals. 
 For both system models, we derive  the channel capacity subject to an average power constraint at the relay and an additional energy transmission cost at the EH source. 
We compare the derived capacities to the achievable rates of several  benchmark schemes. Our results show that using the optimal input distributions at both the EH source  and the relay is essential for high performance.  Moreover,   we demonstrate that neglecting the energy transmission cost at the source can result in a severe overestimation of the achievable performance.

\end{abstract}

\IEEEpeerreviewmaketitle

\newtheorem{theorem}{Theorem}
\newtheorem{corollary}{Corollary}
\newtheorem{remark}{Remark}
\newtheorem{lemma}{Lemma}
\newtheorem{defi}{Definition}
%\thispagestyle{empty}

%%%%%%%%%%%%%%%%%%%%%%%%%%%%%%%%%%%%%%%%%%%%%%%%%%%%%%%%%%%%%%%%%
\section{Introduction}

\begin{figure}
  \centering
  \includegraphics[height=0.25\linewidth]{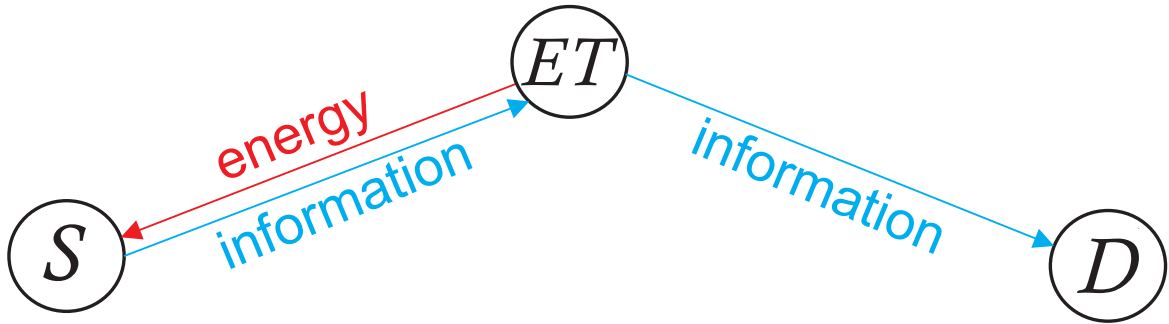}
\caption{System model comprised of an EH source (S), a wireless energy transmitter (ET) acting also as a relay, and a destination (D).}
\label{fig:test}
 \vspace*{-6mm}
\end{figure}

Future  wireless communication devices are expected to be powered by harvesting  freely available ambient energy, such as solar, thermal, and electro-magnetic, and/or  radio frequency (RF) energy transmitted from dedicated wireless energy transmitters \cite{6710085}. Since energy harvesting (EH) from natural resources is usually climate and location dependent, it may not be suitable for small and mobile wireless communication devices. For such devices, wireless energy transfer (WET) from WET transmitters is an appealing solution for providing a perpetual power supply \cite{7081084}. 
However, although WET is a very useful technology,  due to the high path loss attenuation, the RF energy  emitted by the WET transmitter is severely diminished when it is received at   EH  device \cite{7081084}. As a result, EH information sources powered by WET  are constrained to communicate over short distances and with low data rates. One possible solution to overcome this limitation  is to utilize the WET transmitters also as information  relays which forward the information received from the EH  information sources to the  intended  destinations. This is appealing since WET transmitters are expected to have a perpetual energy supply which can  be used to facilitate both WET and  information forwarding (i.e., relaying).  A practical example where this architecture   might be beneficial is the case where  a sensor (i.e., the EH source) is embedded in a concrete wall to monitor the quality of the concrete, and a relay powers up the sensor using WET before relaying the measurement information received from the sensor to  a WiFi access-point. In this paper, we investigate the simplest system model  for such a scenario comprising   an EH  source, an out-of-band\footnote{An out-of-band FD relay receives and transmits signals at the same time but in different frequency bands.} full-duplex (FD) relay  performing WET  and information forwarding for the EH source, and a destination, cf. Fig.~\ref{fig:test}.  %
The FD relay transmits energy to the EH source and information to the destination in one frequency band and receives information from the EH source in another frequency band. 
Thereby, we consider two scenarios for the battery of the EH source node. In the first scenario,   we assume that the EH source is too small to be  equipped with a battery and, as a result, cannot store energy. In this case, the RF energy  harvested during one symbol interval can only be used  in the following symbol interval. In the second scenario, we assume that the EH source is equipped with an ``unlimited battery''\footnote{The term ``unlimited battery'' is used to denote a battery with unlimited storage capacity.} in which it can store the harvested RF energy. In this case, the RF energy  harvested during one symbol interval can  be used  in any of the following symbol intervals.  Moreover, for both scenarios, we assume that a part of the transmit energy is dissipated at the EH source, and only the remaining part can be used for information transmission, i.e., a non-zero energy transmission cost is incurred. For this relay channel, we investigate the channel capacity  for the case when the source-relay and relay-destination channels are both non-fading additive white Gaussian noise (AWGN) channels, and  the relay has an  average power constraint.

Communication systems with EH and WET have recently attracted significant interest, which has led to the study of different types of channel models, including the point-to-point channel, multiple-access channel, broadcast channel, and relay channel  \nocite{6710085, 7081084, 6216430, 6766774, 6190022,  6623062, 6893029, zlatanov_EH, 6108309, 6489506,
  5958560,6702854,6725596, 
 6552840, 6898012, 7403938,7088643, 6970915, 7440849, 7295551} \cite{6710085}-\cite{7295551}.   
In particular,
the capacity of the EH AWGN point-to-point channel\footnote{We note that the point-to-point channel  and the relay channel   differ  significantly from an information-theoretic perspective. As a result, the capacity expressions and the corresponding capacity-achieving coding schemes for  the point-to-point channel are not directly applicable to the relay channel.}, comprised of an EH source equipped with an unlimited battery and a destination,    was derived in \cite{6216430}, \cite{6766774}.    
Whereas, the capacity of the   EH  AWGN point-to-point channel with a batteryless source was derived in \cite{6190022}, where it was shown that the capacity achieving input distribution is discrete and amplitude constrained.  Receiver designs for simultaneous wireless information and energy transfer were proposed in  \cite{6623062}. Achievable rates for the EH AWGN multiple-access      channel (MAC)  were studied in \cite{6893029}, where it was shown that under certain asymptotic conditions  the achievable rate converges to the capacity of the non-EH AWGN MAC. Achievable rates for different types of EH AWGN  broadcast channels were studied in \cite{zlatanov_EH}-\cite{6489506}. On the other hand,   relay channels with EH and WET have been investigated in   \cite{5958560}-\cite{7295551}, where   \cite{5958560}-\cite{6725596} assumed EH from natural sources, \cite{6552840}-\cite{7403938} assumed WET from   source to   relay, and   \cite{7088643}-\cite{7295551} investigated WET   from  relay to   source. 
In particular, in \cite{5958560}, the authors  considered   cooperative EH communications  where, in each time slot, the EH source transmits its data to the destination either  directly or with the help of an EH relay.
The authors in \cite{6702854} investigated the outage probability of a general wireless cooperative network  where multiple pairs of sources and destinations communicate through a single EH relay. Reference \cite{6725596} investigated achievable rates for a buffer-aided relay channel comprised of an EH source, an EH buffer-aided relay, and a destination.
For relay channels with WET  from  source to  relay,  \cite{6552840} studied the corresponding  achievable rates and outage probabilities,  \cite{6898012} derived   throughputs for  instantaneous transmission, delay-constrained transmission, and delay tolerant transmission, and   \cite{7403938} studied throughput maximization for an 
FD multi-antenna relay channel where a time-switching protocol was employed by the   relay.

On the other hand, \cite{7088643}-\cite{7295551} studied   
achievable rates for   relay channels with WET from  relay to   source, which is similar to the relay channel considered in this paper. However,  different from the  study conducted in this paper,  \cite{7088643}-\cite{7295551}   considered   half-duplex relays, did not take into account   the energy transmission cost at the EH source, and only derived achievable rates, not capacities. Moreover, the achievable rates presented  in \cite{7088643}-\cite{7295551}  are achieved with   heuristic protocols which do not provide insight into the capacity of the considered relay channels.

As can be  seen from the discussion above,   an explicit characterization of the capacity and the capacity-achieving coding scheme of the considered relay channel  with WET  is not available in the literature. This motivates the information-theoretic analysis presented in this paper, which offers a more fundamental insight into the   limits of communications for the considered  relay channel. Our information-theoretic analysis reveals that  the capacity is achieved when the transmit signal of the relay is used to simultaneously power the EH source and to convey information to the destination.  In addition, we show that  the EH source has to be silent in a fraction of the symbol intervals to conserve energy. Moreover, for the case when the EH source is equipped with an unlimited battery, we show that  the EH source should use these silent symbol intervals for encoding additional information for the relay. 
As  numerical examples, we compare the derived  capacities to the achievable rates of several  benchmark schemes. The comparison reveals  that using the optimal input distributions at both the EH source and the  relay is essential for achieving high performance.  Moreover,   we illustrate that neglecting the additional energy transmission cost at the EH source can result in a severe overestimation of the achievable performance.

The remainder of this paper is organized as follows. In Section~\ref{sec-sm}, we introduce the system and channel models. In Sections~\ref{sec_capacity} and \ref{sec-4}, we present the capacities of the considered relay channel for the cases of a batteryless EH source and an EH source with an unlimited battery, respectively. In Section~\ref{sec-num}, we provide numerical results. Finally, Section~\ref{sec-conc} concludes the paper.

\section{System and Channel Models}\label{sec-sm}

In the following, we formally introduce the considered   system and channel models.

\subsection{System Model}\label{sec_sm_1}

We consider a two-hop FD relay channel, comprised of an  EH source $S$, an out-of-band FD relay $R$, and a destination $D$, where a direct source-destination link does not exist due to e.g.  the large distance/heavy blockage between the source and the destination, cf. Fig.~\ref{fig:test}. For this relay channel, we assume that the source-relay and relay-destination links are not impaired by fading. Moreover, we assume that the source is an EH node that is powered wirelessly by the RF energy received from the relay, which has  an average transmit power constraint denoted by  $P_R$. Let  $E_{H,i}$ denote the harvested energy at the EH source during symbol interval\footnote{We use the terms ``symbol interval'' and ``channel use'' interchangeably, i.e., during one symbol interval the channel can be used only once for sending one symbol. Hence, one symbol spans  one symbol interval. On the other hand, a codeword is comprised of many symbols and thereby spans many symbol intervals.} $i$. The source uses the harvested RF energy to transmit information to  the relay  which then forwards the received information to the destination. We assume that the EH source can simultaneously harvest energy and transmit information since the simultaneous reception and   transmission occurs in different frequency bands. Similarly, we assume that the WET relay can simultaneously receive information and transmit energy since the simultaneous reception and   transmission occurs in different frequency bands. In addition, for the considered relay channel, we assume that the transmission of a symbol with non-zero energy at the EH source incurs  an   energy transmission cost, denoted by $P_{\rm C}$,   due to dissipation of the input energy  in  the transmitter's   circuitry.
On the other hand, we assume that the EH source can transmit  a zero-energy symbol without any additional cost just by staying silent during a given symbol interval, i.e., without applying any input energy to the transmitter's circuitry\footnote{The adopted constant energy cost model for  non-zero symbols was introduced in \cite{4723323} for conventional (i.e., non-EH)  and in \cite{6766774} for EH communication systems. In general, this energy cost model can be seen as a first order approximation of the actual energy cost  incurred   in communication  systems.}  \cite{6766774, 4723323}.

Furthermore, we consider two cases for the battery of the EH source, namely a batteryless EH source and an  EH source   equipped with an unlimited battery. In the first case,  the EH source cannot  store   the harvested energy. In particular, for the  batteryless EH source, the energy harvested  during symbol interval $i-1$, $E_{H,i-1}$, can be used for transmission \textit{only} in symbol interval $i$. If the energy harvested  during symbol interval $i-1$, $E_{H,i-1}$, is not used during symbol interval $i$, then we assume that the energy $E_{H,i-1}$ is lost and cannot be used in future symbol  intervals. 
 In the second case, the EH source with  unlimited battery can store any amounts of harvested energy   for unlimited amount of time. As a result, the EH source can drain energy from its battery and use it for transmission of information in any future symbol interval.  

\begin{remark}
By performing an information-theoretic analysis for the two extreme cases in terms of the energy storage capacity of  the EH source, namely a batteryless EH source and an EH source with unlimited battery,  we obtain   channel capacities which constitute lower and upper bounds on the channel capacity for the case when the EH source   can store a finite amount of energy.
\end{remark}

\subsection{Channel Model}\label{sec_cm}
The time-discrete memoryless two-hop FD  relay channel is defined by $\mathcal{X}_S$, $\mathcal{X}_R$,     $\bar{\mathcal{Y}}_R$, $\bar{\mathcal{Y}}_D$,   and $p(\bar y_R,\bar y_D|x_S,x_R)$, where $\mathcal{X}_S$ and $\mathcal{X}_R$ are the  input alphabets at  the source and the relay, respectively,    $\mathcal{\bar Y}_R$ and $\mathcal{\bar Y}_D$ are the output alphabets at   the relay and the destination, respectively,   and $p(\bar y_R,\bar y_D| x_S,x_R)$ is the probability distribution   on   $\mathcal{\bar Y}_R\times\mathcal{\bar Y}_D$  for given $x_S\in \mathcal{X}_S$ and $x_R\in \mathcal{X}_R$. In symbol interval $i$, let $X_{S,i}$ and $X_{R,i}$ denote the random variables (RVs) modeling the transmit symbols of the source and the relay, respectively,  and  let    $\bar Y_{R,i}$  and $\bar Y_{D,i}$ denote the RVs modeling the received symbols at the  relay  and destination, respectively. Then, $x_{S,i}\in \mathcal{X}_S$, $x_{R,i}\in \mathcal{X}_R$, $\bar y_{R,i}\in \mathcal{\bar Y}_R$, and $\bar y_{D,i}\in \mathcal{\bar Y}_D$ are the realizations of $X_{S,i}$,  $X_{R,i}$,  $\bar Y_{R,i}$,  and $\bar Y_{D,i}$, respectively.

The considered channel is memoryless in the sense  that given the input symbols   for the $i$-th channel use, the $i$-th output symbols  are independent from all previous input symbols. As a result,    $p(  \bar y_{R}^n, \bar y_{D}^n| x_{S}^n, x_{R}^n)$, where the  notation $a^n$ is   used   to denote the ordered sequence  $a^n=(a_{1}, a_{2},..., a_{n})$, can be factorized as $
     p(  \bar y_{R}^n, \bar y_{D}^n| x_{S}^n, x_{R}^n) =\prod_{i=1}^n p(\bar y_{R,i}, \bar y_{D,i}| x_{S,i}, x_{R,i}).$ Moreover, since the considered two-hop FD  relay channel does not have a direct source-destination link, this relay channel belongs to the class of degraded relay channels defined in \cite{cover}. As a result, $p(\bar y_R,\bar y_D| x_S,x_R)$ can also be written as $p(\bar y_R, \bar y_D| x_S,x_R)=p(\bar y_R| x_S,x_R)p(\bar y_D|x_R)$, where we have used $p(\bar y_D|x_S,x_R,\bar y_R)=p(\bar y_D|x_R)$.

Let $w\in \{1, 2, . . . , W\}$ be the message that the EH source wants to transmit to the destination. Let us define the  encoding function for $w$ at the source for channel use $i$   as
\begin{align}\label{eq_bb1}
x_{S,i} = 
\left\{
\begin{array}{ll}
g_{S,i}(w, E_{H,i-1},P_{\rm C}) &\textrm{ for the batteryless EH source}\\
g_{S,i}(w, E_H^{i-1},P_{\rm C})& \textrm{ for the unlimited battery EH source.}
\end{array}
\right.
\end{align}
In (\ref{eq_bb1}), $x_{S,i}$ is the output of the source's encoder, whereas message $w$, harvested energy $E_{H,i-1}$  or the sequence of harvested energies $E_H^{i-1}$, and energy transmission cost $P_{\rm C}$ are the inputs at the source's encoder. On the other hand,  the   encoding function at the relay for channel use $i$ is defined as $
x_{R,i} = g_{R,i}(\bar Y_{R}^{i-1}),
$
where $x_{R,i}$ is the output of the relay's encoder for channel use $i$ and the sequence $\bar Y_{R}^{i-1}$ is the input of the relay's encoder. Finally,   the   decoding function at the destination   is defined as
$
\hat w = g_D(\bar Y_{D}^{n}),
$
where $n$ denotes the total number of channel uses, $\hat w$ is an estimate of   transmitted message $w$, and $\bar Y_{D}^{n}$ is the input sequence of the decoder at the destination.

Here, we assume that the source-relay  and relay-destination links are  real-valued AWGN  channels with constant channel gains $h_{SR}$ and $h_{RD}$, respectively, and noise variances    $\bar \sigma_R^2$  and  $\bar \sigma_D^2$, respectively\footnote{Similar to \cite{cover}, as a first step for investigating the capacity of the considered relay channel,  we do not consider fading and assume   real-valued channel inputs and outputs. Deriving the channel capacity when fading is present is a much more 
difficult task since the source-relay and relay-destination channel gains vary from one channel use to the next. As a result, the fading case is left for future investigation. Similarly, deriving the capacity for the multi-antenna case, though of high interest, is beyond the scope of this paper.}.  Since the channel gains $h_{SR}$ and $h_{RD}$  are assumed to be constant, i.e., they do not vary with time,   given sufficient time, the channel gains $h_{SR}$ and $h_{RD}$ can be estimated almost perfectly using pilot symbols, see \cite{841172}.
Hence,    we assume that $h_{SR}$ and $h_{RD}$ are perfectly known at all three nodes and are fixed during the entire transmission.
In symbol interval $i$, let     $\bar Z_{R,i}$  and $\bar Z_{D,i}$ denote the RVs modeling the AWGN at    relay  and destination, respectively.   Consequently, the 
RVs modeling the received symbols at relay and destination in channel use $i$, $\bar Y_{R,i}$ and $\bar Y_{D,i}$,  are given by
\begin{align}\label{eq_io_1}
\bar Y_{R,i} = h_{SR}  X_{S,i} + \bar Z_{R,i}\; \textrm{ and }\;
\bar Y_{D,i} = h_{RD}  X_{R,i} + \bar Z_{D,i}.
\end{align}
For notational simplicity and without loss of generality, instead of studying the capacity for the  input-output model  in  (\ref{eq_io_1}),  we can  study instead the capacity  using  the following input-output model
\begin{align}\label{eq_io_2}
  Y_{R,i} =   X_{S,i} +   Z_{R,i} \textrm{ and } Y_{D,i}=   X_{R,i} +  Z_{D,i}, 
\end{align}
where 
\begin{align}
 Y_{R,i} =\frac{\bar Y_{R,i}}{h_{SR}}, \;\;
Y_{D,i}=\frac{\bar Y_{D,i}}{h_{RD}}, \;\;
Z_{R,i} =\frac{\bar Z_{R,i}}{h_{SR}}, \;\; \textrm{ and } \;\;
Z_{D,i}=\frac{\bar Z_{D,i}}{h_{RD}}. \label{eq_io_3b}
\end{align}
Using (\ref{eq_io_3b}), we can obtain the variances of the equivalent noises  at   relay and  destination, $Z_{R,i}$ and $Z_{D,i}$,   as 
$\sigma_R^2=  \bar \sigma_R^2/h_{SR}^2$  and $\sigma_D^2=  \bar \sigma_D^2/h_{RD}^2$, 
respectively, where $\bar \sigma_R^2$  and  $\bar \sigma_D^2$ are the variances of the AWGNs $\bar Z_{R,i}$ and $\bar Z_{D,i}$, respectively.

\subsection{Energy Harvesting Model}
We assume that the harvested energy at the EH source in symbol interval $i$, $E_{H,i}$,   is a deterministic function of the transmit symbol at the relay in symbol interval $i$,  $X_{R,i}$, which we write explicitly as  $E_{H,i}=E_H(X_{R,i})$, where $E_H(X_{R,i})$ is a deterministic function of $X_{R,i}$. Since $E_H(X_{R,i})$ represents the harvested energy, we assume that $E_H(X_{R,i})$ is a non-decreasing function of $X_{R,i}^2$.  For the presented analysis, we do not need to impose any additional constraints on  $E_H(X_{R,i})$.  Hence, the  model adopted for the energy harvested at the EH source during symbol interval $i$ is very general. 
One example for $E_H(X_{R,i})$, which was considered in numerous  previous works    \cite{6216430}-\cite{7295551} and is included in our general model as a special case, is
\begin{align}\label{eq_E1}
E_H(X_{R,i})= \eta  h_{RS}^2 X_{R,i}^2,
\end{align} 
where $ h_{RS}$ is the relay-source channel gain   and $0<\eta<1$ denotes the energy harvesting efficiency.

Having defined the considered relay channel and the  harvested energy, in the following, we derive the channel capacities  for the two different battery scenarios at the EH source.

\section{Channel Capacity when the Source is Batteryless}\label{sec_capacity}

In the following, we  study the capacity for the case when the EH source is batteryless. For clarity of presentation,  we first provide a summary of the procedure adopted for deriving the capacity.

\subsection{Summary of the Capacity Derivation}
The approach employed for derivation of the capacity of the considered two-hop FD relay channel with WET and batteryless EH source is summarized in the following five steps.

1) The   two-hop FD relay channel with WET and  batteryless EH source is written equivalently as a conventional (i.e., non-EH) two-hop FD relay channel with fading source-relay channel, where the  fading states of the source-relay channel are  known completely at both the source and the relay, and the source has a constraint on the  amplitude of its input symbols.
 
2) Since the equivalent conventional two-hop FD relay channel does not have a direct source-destination link, this relay channel belongs to the class of degraded relay channels as defined in \cite{cover}. Consequently, the converse of the capacity  for the degraded relay channel derived in  \cite{cover}  is also a converse of the capacity for the equivalent conventional two-hop FD relay channel. As a result,  the general capacity expression for the degraded relay channel in \cite{cover} is also the capacity expression for the equivalent conventional two-hop FD relay channel.

3) Next, we simplify the general capacity expression for the degraded relay channel in \cite{cover} exploiting the following  characteristics of the equivalent conventional two-hop FD relay channel: (a) AWGN source-relay and relay-destination channels, (b) fading source-relay channel with fading states  that are known  at   source and   relay, and (c) amplitude constrained inputs at the source.

4) Subsequently, we derive the optimal input distributions at  the source and the relay for the equivalent conventional two-hop FD relay channel and insert   them into the general capacity expression for the degraded relay channel in \cite{cover}. As a result, we obtain  the final capacity expression in Theorem~\ref{theo_1}.

5) For the achievability of the derived capacity, we resort to the general  capacity-achieving  coding scheme for the degraded relay channel presented in  \cite{cover}. This coding scheme requires $N\to\infty$ time slots, where each time slot is comprised of $k\to\infty$ symbol intervals  with $Nk=n$, and  a decode-and-forward (DF) relay. Furthermore, in each time slot, the source and the relay transmit with the minimum of the  capacities of the corresponding source-relay and relay-destination channels. Since the equivalent conventional two-hop FD relay channel has a   fast-fading\footnote{A fast-fading channel is a channel where the fading  gain changes from one symbol interval to the next.} AWGN source-relay channel and full CSI is available at  source and relay,  we use the coding scheme\footnote{The coding scheme in \cite{641562} is a capacity-achieving coding scheme for the fast-fading AWGN channel with full CSI at both the transmitter and the receiver.} in \cite{641562} in order to achieve the  capacity of the source-relay channel  in each time slot.

\subsection{Equivalent Relay Channel}
In this subsection, we model  the considered two-hop FD relay channel with WET and batteryless EH source as an equivalent conventional two-hop FD relay channel with fading source-relay channel, where the fading states are known   at   source and   relay, and an amplitude constrained source.

Since the source is batteryless, the energy  harvested during symbol interval $i-1$, $E_H(X_{R,i-1})$, cannot be stored at the source and can only be used for  transmission of the source  symbol during symbol interval $i$, $X_{S,i}$. On the other hand, since the energy of symbol $X_{S,i}$ is $X_{S,i}^2$, we require that  $X_{S,i}^2\leq E_H(X_{R,i-1})$ has to hold, i.e., the source cannot transmit more energy in symbol interval $i$ than what it has harvested during symbol interval $i-1$. However, because of the additional  energy transmission cost that incurs during transmission, $P_{\rm C}$, the energy of symbol $X_{S,i}$, $X_{S,i}^2$, has to satisfy the following  more stringent constraint
\begin{align}\label{eq_E2}
 X_{S,i}^2 \leq \max\{ 0, E_H(X_{R,i-1}) -P_{\rm C} \}\triangleq f(X_{R,i-1}),
\end{align}
where function $f(X_{R,i-1})$ is introduced for notational simplicity.
Condition (\ref{eq_E2}) means that if the energy harvested  during symbol interval $i-1$, $E_H(X_{R,i-1})$, exceeds the energy transmission cost, $P_{\rm C}$, then the energy of symbol  $X_{S,i}$, $X_{S,i}^2$, can take any value between zero and $E_H(X_{R,i-1}) -P_{\rm C}$, i.e., $0\leq X_{S,i}^2\leq f(X_{R,i-1})$. Otherwise, if the energy harvested  during symbol interval $i-1$, $E_H(X_{R,i-1})$, is smaller than the energy transmission cost, $P_{\rm C}$, then the energy of symbol  $X_{S,i}$, $X_{S,i}^2$, can only be zero, i.e., $X_{S,i}^2=0$. Hence, $f(X_{R,i-1})=\max\{ 0, E_H(X_{R,i-1}) -P_{\rm C} \}$ represents an upper bound on the available transmission energy for symbol $X_{S,i}$.

Since the absolute value of symbol $X_{S,i}$ is uniquely determined by its energy, $X_{S,i}^2$, from (\ref{eq_E2}), we can obtain the limits for the value of   source symbol $X_{S,i}$   as
\begin{align}\label{eq_2}
 -  \sqrt{f(X_{R,i-1})} \leq  X_{S,i}     \leq  \sqrt{f(X_{R,i-1})}  .
\end{align}
From (\ref{eq_2}), we can conclude that if $P_{\rm C}\geq E_H(X_{R,i-1})$, i.e., if  $f(X_{R,i-1})=0$ holds, then the source can \textit{only} transmit the symbol zero (i.e., it can only  be silent) in symbol interval $i$ since the source does not have enough energy  for the transmission of any other symbol. On the other hand, if $P_{\rm C}< E_H(X_{R,i-1})$, i.e., if  $f(X_{R,i-1})>0$ holds, then the source can  transmit any symbol in the range between $ -  \sqrt{f(X_{R,i-1})}$ and  $\sqrt{f(X_{R,i-1})}$. 
Hence, the effect of WET for a batteryless EH source can be modeled by an amplitude constrained input at the EH source, cf.   (\ref{eq_2}). This is also in line with the model and results for the batteryless EH source presented in \cite{6190022}. Hence, the capacity of the considered relay channel with batteryless EH source can be found using  the input-output relations in (\ref{eq_io_2}), where the input at the source, $X_{S,i}$, has to meet the amplitude constraint  in (\ref{eq_2}). In the following, we provide an  equivalent   expression for the amplitude constraint in   (\ref{eq_2}), and thereby obtain an equivalent channel model.
  
In particular,  (\ref{eq_2}) can be represented equivalently as
\begin{align}\label{eq_2a}
   X_{S,i} = V_{S,i} \sqrt{f(X_{R,i-1})},
\end{align}
 where $ V_{S,i} $  is an RV, which satisfies
\vspace*{-5mm}
\begin{align}\label{eq_2b}
-1\leq V_{S,i}\leq 1.
\end{align}
Eqns. (\ref{eq_2a}) and (\ref{eq_2b}) provide  a different perspective on how source  symbol $X_{S,i}$ can be generated for channel use $i$. In particular,  the source can first generate the symbol  $V_{S,i}$, which can take any value between $-1$ and $1$, and then multiply $V_{S,i}$ with the coefficient  $\sqrt{f(X_{R,i-1})}$ in order to produce $X_{S,i}$. Inserting (\ref{eq_2a}) into   (\ref{eq_io_2}), we obtain 
%\vspace*{-12mm}
\begin{align}\label{eq_io_2a}
  Y_{R,i} =   V_{S,i} \sqrt{f(X_{R,i-1})}  +   Z_{R,i},\textrm{ where } -1\leq V_{S,i}\leq 1, \textrm{ and } Y_{D,i}=   X_{R,i} +  Z_{D,i}.
\end{align}
Hence, instead of using  (\ref{eq_io_2}) and constraint (\ref{eq_2}) for deriving the capacity, we can equivalently use    (\ref{eq_io_2a}). On the other hand, the input-output relation for the source-relay channel  in (\ref{eq_io_2a}) resembles a fast-fading AWGN channel, where the  source imposes an amplitude constraint on its transmit symbols, $V_{S,i}$, as in    (\ref{eq_2b}),  and where the ``fading channel gain'' in channel use $i$ is $\sqrt{f(X_{R,i-1})}$, which is known at source and relay, i.e.,  the  source and the relay  have full CSI\footnote{Since the function $f(X_{R,i-1})$  is a deterministic function of the relay's transmit symbol $X_{R,i-1}$, it is clear that the relay  knows the value of $f(X_{R,i-1})$ even before the start of  symbol interval $i$. On the other hand, since the source knows how much energy it harvested during symbol interval $i-1$, it can calculate $f(X_{R,i-1})$ in symbol interval $i$ using (\ref{eq_E2}).}.

 \begin{figure}
\centering
\includegraphics[width=5in]{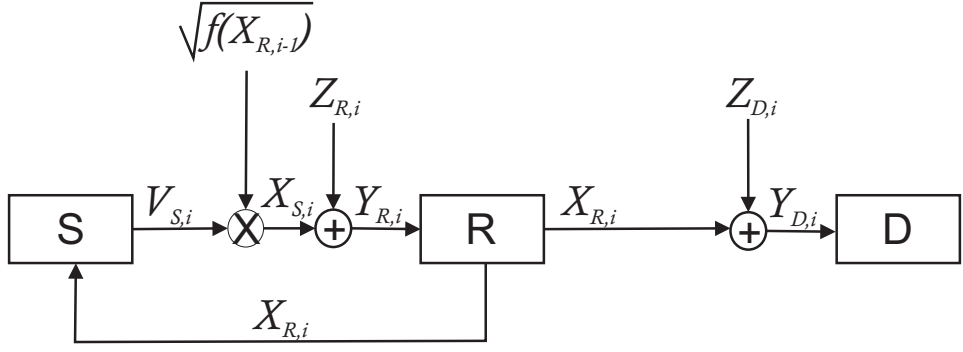}
\vspace*{-3mm}
\caption{An equivalent model of the considered relay channel, where $V_{S,i}$ is constrained as in (\ref{eq_2b}).}  \label{fig_2}
\vspace*{-6mm}
\end{figure}

Hence,  the considered two-hop FD relay channel with WET and batteryless EH source can equivalently be represented as a conventional two-hop FD relay channel with fast-fading on the source-relay channel, where the fading states  are known  at  source and   relay, and amplitude constrained inputs at the source, cf. (\ref{eq_2b}). This equivalent   relay channel is illustrated in Fig.~\ref{fig_2}.  Since the two relay channels are equivalent,   the capacities of the two relay channels  are identical. Hence, by deriving the capacity   of the relay channel shown in Fig.~\ref{fig_2}, we also obtain the capacity   of the original two-hop FD relay channel with WET and batteryless EH source. 

A general expression for the  capacity of the considered relay channel with WET and batteryless EH source  is given in the following lemma.
\begin{lemma}\label{lemma_1a}
The capacity of the two-hop FD relay channel with batteryless EH source is given by the following general expression
\begin{align}\label{eq_5}
C=\max_{p(x_R)\in \mathcal{P}}  \min \Bigg\{&\sum\limits_{x_R\in \mathcal{X}_R} \max_{p(v_S|x_R)\in \mathcal{P}} I(V_S;Y_R|X_R=x_R) p(x_R)\;  ; \; I(X_R;Y_D) \Bigg\} \nonumber\\
\textrm{Subject to } &\mathrm{C1:}  -1 \leq  V_S   \leq 1 , \nonumber\\
&\mathrm{C2:}   \sum_{x_R\in \mathcal{X}_R} x_R^2 p( x_R) \leq P_R ,
\end{align}
where $\mathcal{P}$ is the set of all possible  probability distributions, $P_R$ is the average transmit power constraint at the relay, $p(v_S|x_R)$ is the conditional probability distribution of $V_S$ given $X_R$, and $p(x_R)$ is the probability distribution of $X_R$. 
\end{lemma}
\begin{IEEEproof}
Please refer to Appendix~\ref{app_1a}.
\end{IEEEproof}
From Lemma~\ref{lemma_1a}, we see that in order to obtain a more  explicit expression for the capacity of the considered relay channel with a batteryless EH source, we need to obtain the optimal input distributions $p(v_S|x_R)$ and $p(x_R)$, denoted by  $p^*(v_S|x_R)$ and $p^*(x_R)$, respectively, as the solutions of (\ref{eq_5}). These distributions are provided in the following two subsections.

\subsection{Optimal $p^*(v_S| x_R)$}\label{sec_pv}
The optimal input distribution at the EH source, $p^*(v_S| x_R)$, and the corresponding  expression for the mutual information $ \max\limits_{p(v_S|x_R)\in \mathcal{P}}I(V_S;Y_R|X_R=x_R)$ are provided in the following lemma.  To this end, let us  first define   function $\delta(\cdot)$ as follows:  $\delta(x)=0$ if $x\neq 0$ and $\delta(x)=1$ if $x= 0$.
\begin{lemma}\label{lemma_2a}
The optimal input distribution at the EH source, $p^*(v_S| x_R)$  is discrete with a finite number of probability mass points, and can be written in a general form as
\begin{align}\label{eq_pv}
p^*(v_S|x_R)=\sum_{k=1}^{K^*(x_R)} p^*_{S,k}(x_R)   \delta \big(v_S - v^*_k(x_R) \big), \quad  -1\leq v_k^*(x_R)\leq 1 , \;\forall k,
\end{align}
where  $v^*_k(x_R) $ are the optimal values that  $v_S$ can assume,  $p^*_{S,k}(x_R) ={\rm Pr}\{V_S=v_k(x_R)|X_R=x_R\}$, where ${\rm Pr}\{\cdot|\cdot\}$ denotes conditional probability, is the conditional  probability of the outcome $V_S=v^*_k(x_R)$, and $K^*(x_R)$ is the optimal number of possible values that $v_S$ can assume.    
The corresponding expression for the mutual information $ \max\limits_{p(v_S|x_R)\in \mathcal{P}}I(V_S;Y_R|X_R=x_R)$ is given by
\begin{align}\label{eq_Is}
 & \max\limits_{p(v_S|x_R)\in \mathcal{P}} I(V_S;Y_R|X_R=x_R) =   - \int_{-\infty}^\infty \sum_{k=1}^{K^*(x_R)}  \frac{p^*_{S,k}(x_R)}{\sqrt{2\pi \sigma_R^2}} \exp\left( -\frac{\left(y_R-v_k^*(x_R) \sqrt{f(x_R)} \right)^2}{2 \sigma_R^2} \right) \nonumber\\
&\times \log_2\left(   \sum_{k=1}^{K^*(x_R)}  \frac{p^*_{S,k}(x_R)}{\sqrt{2\pi \sigma_R^2}} \exp\left( -\frac{\left(y_R-v^*_k(x_R) \sqrt{f(x_R)} \right)^2}{2 \sigma_R^2} \right)   \right) d y_R - \frac{1}{2} \log_2\big(2\pi e \sigma_R^2\big). 
\end{align}
\end{lemma}
\begin{IEEEproof}
Please refer to Appendix~\ref{app_1b}.
\end{IEEEproof}

There are  special cases for which  the optimal input distribution 
   $p^*(v_S|x_R)$ and the corresponding maximal mutual information $\max\limits_{p(v_S|x_R)\in \mathcal{P}} I(V_S;Y_R|X_R=x_R)$ are known in closed form. In particular,  $p^*(v_S|x_R)$ and  $\max\limits_{p(v_S|x_R)\in \mathcal{P}} I(V_S;Y_R|X_R=x_R)$  are known in closed form if  $f(x_R) /\sigma^2_R$ is sufficiently small and large\footnote{Parameter  $f(x_R) /\sigma^2_R$ can be interpreted as the normalized transmit power at the EH source for $X_R=x_R$, i.e., the normalized transmit power at the EH source for a given symbol interval for which $X_R=x_R$ holds.}, respectively.  These closed-form expressions are provided in the following corollary.

\begin{corollary}\label{corollary_1c}
In the special case when  $f(x_R) /\sigma^2_R\ll 1$, 
the optimal input distribution at the source $p^*(v_S|x_R)$  is the binary distribution   
\begin{align}\label{eq_e1}
p^*(v_S|x_R)  = \frac{1}{2} \delta(v_S-1)+ \frac{1}{2} \delta(v_S+1). 
\end{align}
Consequently, the corresponding $ \max\limits_{p(v_S|x_R)} I(V_S;Y_R|X_R=x_R)$ is  given by   
\begin{equation}\label{eq_cap_4a}
 \max_{p(v_S|x_R)} I(V_S;Y_R|X_R=x_R)   = \frac{f(x_R)}{\ln(2)\sigma_R^2}   -  \int_{-\infty}^\infty    \frac{ e^{-t^2/2}}{  \sqrt{2\pi}} \log_2\left(\cosh\left( \frac{f(x_R)}{\sigma_R^2} +\sqrt{\frac{f(x_R)}{\sigma_R^2}} t \right) \right) dt.
\end{equation}
On the other hand, when $f(x_R)/\sigma_R^2\gg 1$,   the optimal input distribution at the source $p^*(v_S|x_R)$ is uniformly distributed with a very  large number of probability mass points in the interval $[-1,1]$. As a result, the optimal input distribution can be accurately approximated by the continuous  uniform distribution  given by
\begin{align}\label{eq_pv_uniform}
p^*(v_S|x_R)= \left\{
\begin{array}{ll}
1/2 &\textrm{if } -1\leq v_S\leq 1\\
0 & \textrm{otherwise}.
\end{array}
\right.
\end{align}
Consequently, the corresponding $ \max\limits_{p(v_S|x_R)} I(V_S;Y_R|X_R=x_R)$ is  given by   
\begin{align}\label{eq_eq1}
\max_{p(v_S|x_R)\in \mathcal{P}} I(V_S;Y_R|X_R=x_R) = \frac{1}{2}\log_2\left( 1+\frac{2 f(x_R)}{\pi e \sigma_R^2} \right).
\end{align}
\end{corollary}
\begin{IEEEproof}
Proofs that the optimal input distribution for low and high SNRs for an amplitude constrained  AWGN channel, i.e., when  $f(x_R) /\sigma^2_R\ll 1$ and $f(x_R) /\sigma^2_R\gg 1$ hold, are given by (\ref{eq_e1}) and   (\ref{eq_pv_uniform}), respectively, are given in \cite{smith1971information} and \cite{jog2015geometric}. Consequently, $ \max\limits_{p(v_S|x_R)} I(V_S;Y_R|X_R=x_R)$ for input distributions given by (\ref{eq_e1}) and   (\ref{eq_pv_uniform}) are derived in \cite[pp. 145]{benedetto1999principles}  and  \cite{smith1971information} as (\ref{eq_cap_4a}) and (\ref{eq_eq1}), respectively.
\end{IEEEproof}

\begin{remark}\label{remark_3}
 As can be seen from   (\ref{eq_pv_uniform}), when $f(x_R)/\sigma_R^2\gg 1$ holds, the optimal input distribution at the source  $p^*(v_S|x_R)$ becomes independent of $x_R$, i.e., independent of the ``fading gain" $f(x_R)$. This property of $p^*(v_S|x_R)$ can be exploited for designing a simple achievability coding scheme, cf.~Remark~\ref{remark_f}.  
\end{remark}

The optimal input distribution at the relay,  $p^*(x_R)$, is provided in the following.

\subsection{Optimal $p^*(x_R)$ and   Capacity Expressions}\label{sec_pr}
In order to obtain the capacity for the considered relay channel with a batteryless EH source,  we first need an expression for $I(X_R;Y_D)$ for a general discrete probability distribution $ 
p(x_R)=\sum_{m=1}^M p_{R,m} \delta(x_R-x_{R,m})$, which is given by \cite{cover2012elements}
\begin{align}\label{eq_I(X_R;Y_D)}
I(X_R;Y_D)= -&\sum_{m=1}^M \int_{-\infty}^\infty  \frac{p_{R,m}}{\sqrt{2\pi \sigma_D^2}}  \exp\left( -\frac{\left(y_R-   x_{R,m} \right)^2}{2 \sigma_D^2} \right)   \nonumber\\
&\quad\times \log_2\left( \sum_{m=1}^M  \frac{p_{R,m}}{\sqrt{2\pi \sigma_D^2}}  \exp\left( -\frac{\left(y_R-   x_{R,m} \right)^2}{2 \sigma_D^2} \right)     \right) d y_R  - \frac{1}{2} \log_2\big(2\pi e \sigma_D^2\big).
\end{align}
Using (\ref{eq_I(X_R;Y_D)}),  the optimal $p^*(x_R)$ and the  capacity  are provided in the following theorem. 
\begin{theorem}\label{theo_1}
There are three cases for the optimal input distribution of the relay,  $p^*(x_R)$,  and the capacity of the considered relay channel with a batteryless EH source.

Case 1: Let $p^+(x_R)$ be the optimal solution of
\begin{align}\label{eq_cap_1}
C= \max_{p(x_R)\in \mathcal{P}} & \sum\limits_{x_R\in \mathcal{X}_R} \max_{p(v_S|x_R)} I(V_S;Y_R|X_R=x_R) p(x_R) \nonumber\\
\textrm{Subject to }  \mathrm{C1:}  &\sum_{x_R\in \mathcal{X}_R} x_R^2 p( x_R) \leq P_R,
\end{align}
where $\max\limits_{p(v_S|x_R)\in \mathcal{P}}I(V_S;Y_R|X_R=x_R) $ is given in (\ref{eq_Is}). Then, if condition
\begin{align}\label{eq_sdffg}
 C\leq I(X_R;Y_D)\Big|_{p(x_R)=p^+(x_R)}  
\end{align}
holds, where  $I(X_R;Y_D)$ is given in (\ref{eq_I(X_R;Y_D)}), the capacity is given by $C$ in (\ref{eq_cap_1})
 and the optimal $p^*(x_R)$ is $p^+(x_R)$ found as the solution of (\ref{eq_cap_1}).
In this case, the source-relay channel is the performance bottleneck. In particular, even if the relay transmits using   the  distribution which  maximizes the average mutual information of the source-relay channel, the  average mutual information of the source-relay channel is still smaller then or equal to the average mutual information of the relay-destination channel.
 
Case 2: If condition
\begin{align}\label{eq_cap_2}
  C=  \frac{1}{2} \log_2\left(1 + \frac{P_R}{\sigma_D^2} \right) < \int_{-\infty}^{\infty}\max_{p(v_S|x_R)\in \mathcal{P}} I(V_S;Y_R|X_R=x_R) \frac{1 }{\sqrt{2\pi P_R}} \exp\left(-\frac{x_R^2}{2P_R}\right) d x_R     
\end{align}
holds, where $\max\limits_{p(v_S|x_R)\in \mathcal{P}}I(V_S;Y_R|X_R=x_R) $ is given in (\ref{eq_Is}), then the capacity is given by $C$ in (\ref{eq_cap_2}) and the optimal $p^*(x_R)$ is the zero-mean Gaussian distribution with variance $P_R$. In this case, the relay-destination channel is the bottleneck. In particular, even if the relay transmits  Gaussian distributed symbols  with which it achieves the capacity of the relay-destination channel, the rate of the  relay-destination channel is still  smaller than the rate of the source-relay channel.
 Hence, in this case, the capacity is equal to the capacity of the relay-destination link, and therefore, even equipping a   battery at the source would not increase the capacity.

Case 3: If both  conditions (\ref{eq_sdffg}) and (\ref{eq_cap_2}) do not hold, the capacity  is given by 
\begin{align} \label{eq_cap_3}
C= \max_{p(x_R)\in \mathcal{P}} & I(X_R;Y_D)\nonumber \\
\textrm{Subject to }  \mathrm{C1:}  &\sum_{x_R\in \mathcal{X}_R} x_R^2 p( x_R) \leq P_R \nonumber\\
\mathrm{C2:}  & \sum\limits_{x_R\in \mathcal{X}_R} \max_{p(v_S|x_R)\in\mathcal{P}} I(V_S;Y_R|X_R=x_R) p(x_R) =  I(X_R;Y_D),
\end{align}
where $\max\limits_{p(v_S|x_R)\in\mathcal{P}}I(V_S;Y_R|X_R=x_R) $ and $I(X_R;Y_D)$  are given in (\ref{eq_Is}) and (\ref{eq_I(X_R;Y_D)}), respectively. The optimal $p^*(x_R)$  is discrete and  found as the solution of (\ref{eq_cap_3}).

\end{theorem} 
\begin{IEEEproof}
Please refer to Appendix~\ref{app_1}.
\end{IEEEproof}

\begin{remark}\label{rem_2}
The capacity and the optimal distribution   $p^*(x_R)$ for Cases 1 and 3 in Theorem~\ref{theo_1}    can be obtained numerically using a numerical optimization software such as Mathematica. In particular, the optimization problem in  (\ref{eq_cap_1}) is a linear optimization problem and can be easily solved using  numerical optimization software. On the other hand, using its epigraph form, the optimization problem in (\ref{eq_cap_3}) can be written equivalently  as a concave optimization problem  
 as shown in   Appendix~\ref{app_1}    in  (\ref{MPR1}). The equivalent concave optimization problem can then be   solved using   numerical optimization software.   
\end{remark}

In the following, we provide  useful corollaries   for the case when  $f(x_R)/\sigma_R^2\gg 1$  holds $\forall x_R\in \mathcal{X}_R$ for which $f(x_R)\neq 0$.

\begin{corollary}\label{cor_1}
If $f(x_R)/\sigma_R^2\gg 1$  holds $\forall x_R\in \mathcal{X}_R$ for which $f(x_R)\neq 0$, then in the capacity expressions in Theorem~\ref{theo_1}, $\max\limits_{p(v_S|x_R)\in\mathcal{P}} I(V_S;Y_R|X_R=x_R)$ can be replaced by the expression in (\ref{eq_eq1}). 
\end{corollary}
 \begin{IEEEproof}
When $f(x_R)/\sigma_R^2\gg 1$  holds for a given $x_R\in \mathcal{X}_R$, then   $\max\limits_{p(v_S|x_R)\in\mathcal{P}} I(V_S;Y_R|X_R=x_R)$ in (\ref{eq_Is}) converges to the expression in (\ref{eq_eq1}). Otherwise, when   $f(x_R) =0$, then $\max\limits_{p(v_S|x_R)\in\mathcal{P}}I(V_S;Y_R|X_R=x_R)=0$.
\end{IEEEproof}

Corollary~\ref{cor_1} significantly simplifies the expressions in Theorem~\ref{theo_1}.

Since $E_H(x_R)$  given by (\ref{eq_E1}) is a frequently used energy harvesting model, in the following, we derive the capacity for $E_H(x_R)$  given by (\ref{eq_E1}). 

\begin{corollary}\label{cor_2}
If $f(x_R)/\sigma_R^2\gg 1$  holds $\forall x_R\in \mathcal{X}_R$ for which $f(x_R)\neq 0$ and if $E_H(x_R)$ is given by (\ref{eq_E1}), then we have the following three cases.

Case 1: If conditions 
 $\eta  h_{RS}^2 P_{R}> P_{\rm C}$    and  
\begin{align}\label{eq_bj}
C= \frac{1}{2} \log_2\left( 1+\frac{2 (\eta h_{RS}^2 P_R -P_{\rm C})}{\pi e \sigma_R^2} \right) < \frac{P_R}{\ln(2)\sigma_D^2}   -  \int_{-\infty}^\infty    \frac{ e^{-t^2/2}}{  \sqrt{2\pi}} \log_2\left(\cosh\left( \frac{P_R}{\sigma_D^2} +\sqrt{\frac{P_R}{\sigma_D^2}} t \right) \right) dt 
\end{align}
hold, then the capacity is given by $C$ in (\ref{eq_bj}) and the optimal input distribution at the relay is
\begin{align}\label{eq_dfs1}
p^*(x_R)=\frac{1}{2}\delta\left(x_R-\sqrt{P_R}\right)+\frac{1}{2}\delta\left(x_R+\sqrt{P_R}\right).
\end{align}

Case 2: If conditions
 $\eta  h_{RS}^2 P_{R}\leq  P_{\rm C}$  and  
\begin{align}\label{eq_bj2}
&C = \frac{1}{2} \log_2\left( 1+\frac{2 (\eta h_{RS}^2 x_0^2 -P_{\rm C})}{\pi e \sigma_R^2} \right) \frac{P_R}{ x_0^2} \nonumber\\
&<  - \int_{-\infty}^\infty \frac{1}{\sqrt{2\pi \sigma_D^2}} \Bigg\{  \frac{P_R}{ 2x_0^2}  \exp\left( -\frac{\left(y_R-   x_0 \right)^2}{2 \sigma_D^2} \right) +     \frac{P_R}{ 2x_0^2}  \exp\left( -\frac{\left(y_R +  x_0 \right)^2}{2 \sigma_D^2} \right)   \nonumber\\
 &\qquad   +  \left(1-\frac{P_R}{x_0^2}\right) \exp\left( -\frac{ y_R^2}{2 \sigma_D^2} \right)
\Bigg\}  \times \log_2\Bigg[ \frac{1}{\sqrt{2\pi \sigma_D^2}} \Bigg\{  \frac{P_R}{ 2x_0^2}  \exp\left( -\frac{\left(y_R-   x_0 \right)^2}{2 \sigma_D^2} \right) +   \nonumber\\
 &\qquad \frac{P_R}{ 2x_0^2}  \exp\left( -\frac{\left(y_R +  x_0 \right)^2}{2 \sigma_D^2} \right)  +  \left(1-\frac{P_R}{x_0^2}\right) \exp\left( -\frac{ y_R^2}{2 \sigma_D^2}     \right)\Bigg\} \Bigg]  d y_R   - \frac{1}{2} \log_2\big(2\pi e \sigma_D^2\big) 
\end{align}
hold, then the capacity is given by $C$ in (\ref{eq_bj2}) and the optimal input distribution at the relay is
\begin{align}\label{eq_dfs}
p^*(x_R)=\frac{P_R}{2x_0^2} \delta(x_R-x_0) + \left(1-\frac{P_R}{x_0^2}\right) \delta(x_R) + \frac{P_R}{2x_0^2} \delta(x_R+x_0),
\end{align}
 where
\begin{align}\label{eq_sde9a}
x_0= \sqrt{\frac{2P_{\rm C}-e\pi\sigma_R^2}{2\eta h_{RS}^2} - \frac{1}{2\lambda \ln(2)} W\left( -\frac{e\pi\lambda\sigma_R^2\ln(2) }{\eta h_{RS}^2} 2^{\frac{\lambda}{\eta h_{RS}^2}(2P_{\rm C}-e\pi\sigma_R^2)} \right)}.
\end{align}
In (\ref{eq_sde9a}), $W(\cdot)$ is the Lambert W function and  $\lambda$ is given by  
\begin{align}\label{eq_sde10a}
\lambda= \frac{\eta h_{RS}^2}{(2P_{\rm C}-e\pi\sigma_R^2)\ln(2)} W\left( \frac{2P_{\rm C}-e\pi\sigma_R^2 }{e^2\pi\sigma_R^2 } \right).
\end{align}
Case 3: If (\ref{eq_bj}) and (\ref{eq_bj2}) do not hold, the capacity is given by Case 2 or Case 3 in Theorem~\ref{theo_1} with $\max\limits_{p(v_S|x_R)\in\mathcal{P}} I(V_S;Y_R|X_R=x_R)$ and $E_H(x_R)$   given by (\ref{eq_eq1}) and (\ref{eq_E1}), respectively.
\end{corollary}
 \begin{IEEEproof}
Please refer to Appendix~\ref{app_2}.
\end{IEEEproof}

\begin{remark}
For Cases 1 and 2 in Corollary~\ref{cor_2}, we have closed-form expressions for the  capacity. Moreover,  the input distribution at the relay is either  binary shift phase keying (BPSK) (for Case 1) or a simple three-point constellation (for Case 2), which can be considered as a BPSK with an additional zero symbol in the constellation.
\end{remark}

\begin{remark}\label{rem_3a}
We note that the capacity of the considered two-hop FD relay channel with batteryless EH source given in Theorem~\ref{theo_1}, and the  capacity of a conventional (non-EH)  AWGN  two-hop FD relay channel given in \cite{cover}  are very  different. The differences are  due to the amplitude constraint and the energy transmission cost at the EH source,  which are not present for the conventional   AWGN  two-hop FD relay channel  in \cite{cover}.
\end{remark}

In the following, we discuss the achievability of the capacity for the considered relay channel with batteryless EH source  given in Theorem~\ref{theo_1}.

\subsection{Achievability of the Capacity}\label{cap_ach}
Since the considered two-hop FD relay channel belongs to the class of degraded relay channels defined in  \cite{cover}, its capacity can be achieved by the capacity-achieving coding scheme for the degraded relay channel in  \cite{cover}. The capacity-achieving coding scheme in \cite{cover} requires the transmission to be carried out over $N+1$ time slots, where  during each time slot the channel is used $k$ times, where $N \to\infty$ and $ k\to\infty$. Moreover, during each time slot, the source and the relay transmit with  rates which are smaller but arbitrary close to the capacity $C$ given in Theorem~\ref{theo_1}. On the other hand, since the source-relay channel can be modeled as a fast-fading AWGN channel with full CSI at the source and the relay, the  rate $C$  on the  source-relay channel can be achieved using the capacity-achieving coding scheme for a fast-fading AWGN channel with full CSI   proposed in \cite{641562}.

 The coding scheme in \cite{641562} requires the source to use  $F$ codebooks, where $F$ is the number of non-zero ``fading states'' of the source-relay channel. For the considered source-relay channel,  the number of   non-zero ``fading states''   is equal to the number of non-zero values that $f(x_{R})$ can assume. As a result, $F=|f(\mathcal{ X}_{R})|-1$, where set $\mathcal{ X}_{R}$ contains all possible values that $x_R$ can assume and $|\cdot|$ denotes the cardinality of a set. Each of the $F$ codebooks is mapped to a specific non-zero  ``fading state" $f(x_{R})\neq 0$. The codebook corresponding to  ``fading state" $f(x_{R})\neq 0$ contains $2^{kp^*(x_R)R_S(x_R)}$ codewords comprised of $kp^*(x_R)$ symbols where each symbol is generated independently using the distribution  $p^*(v_S|x_R)$, and where $R_S(x_R)$ is given by
\begin{align}\label{eq_r1}
R_S(x_R)= Q\max_{p(v_S|x_R)\in\mathcal{P}} I(V_S;Y_R|X_R=x_R)-\epsilon.
\end{align}
In (\ref{eq_r1}), $\epsilon$ is an arbitrarily small positive number  and $Q$ is a scaling factor given by
\begin{align}\label{eq_Q}
Q =\frac{C}{\sum\limits_{x_R\in\mathcal{X}_R} \max\limits_{p(v_S|x_R)\in\mathcal{P}} I(V_S;Y_R|X_R=x_R) p^*(x_R)},
\end{align}
where $C$ is the capacity given in Theorem~\ref{theo_1}.
The scaling factor $Q$ scales the rate  $R_S(x_R)$ such that the average of $R_S(x_R)$ with respect to $p^*(x_R)$ does not exceed the capacity $C$, i.e., the following holds $
\sum_{x_R\in\mathcal{X}_R} R_S(x_R)  p^*(x_R) <C.
$
On the other hand, the relay uses only one codebook which contains $2^{kR_R}$ codewords comprised of $k$ symbols  where each symbol is generated independently using the distribution  $p^*(x_R)$. The rate of the relay  $R_R$ is set to
$
R_R = C-\epsilon,$
where $C$ is the capacity given in Theorem~\ref{theo_1}. Note that   $R_S(x_R)$ and $R_R$ are related as $ 
R_R= \sum_{x_R\in\mathcal{X}_R} R_S(x_R)  p^*(x_R)=C-\epsilon.
$
Having defined  the above codebooks and rates, in the following, we discuss the transmission of a single message from the source via the relay to the destination in $N+1$ time slots. 

 The source wants to transmit message $w$  selected uniformly from the set $\{1,2,...,2^{n R_R}\}$, which carries $nR_R$ bits of information. This message is split into $N$ smaller messages, denoted by $w(1), w(2),...,w(N)$, where each smaller message carries $k   R_R$ bits and $n=kN$. Each of these smaller messages is transmitted in   a different time slot. Thereby, in the first time slot, the source transmits to the relay message $w(1)$ whereas the relay transmits a known ``dummy'' codeword\footnote{This can be any codeword from relay's codebook and this codeword should be revealed to the source before the start of   transmission.}, which does not carry any information and is used  for powering up the source in the first time slot.   In time slot $b$, where $2\leq b \leq N$, the source transmits   message $w(b)$ to the relay, whereas the relay receives and retransmits to the destination   message $w(b-1)$, which it received from the source in the previous, i.e., $(b-1)$-th time slot. Finally, in the last, i.e., the $(N+1)$-th time slot, the source is silent since it has transmitted all of its messages  and the relay retransmits to the destination  message $w(N)$, which it received  in the previous, i.e., the $N$-th time slot. In the following, we explain how the source and the relay transmit the corresponding  messages in a specific time slot $b$.

The relay transmits   message $w(b-1)$ to the destination in time slot  $b$ by  mapping   message $w(b-1)$ to the corresponding codeword from its codebook and transmitting this codeword to the destination during $k$ symbol intervals. On the other hand, by employing the coding scheme in \cite{641562}, in order for the source to   transmit  message $w(b)$ to the relay in  time slot $b$, the source  splits  message $w(b)$ into $F$ smaller  messages, where each smaller message is mapped to a different ``fading state'' $f(x_R)\neq 0$ and  where the message corresponding to  ``fading state''  $f(x_R)$ carries $k p^*(x_R) R_S(x_R)$ bits of information. 
Now, when ``fading state'' $f(x_R)\neq 0$ occurs in symbol interval $i$ of time slot $b$, the source selects its next untransmitted symbol from the codeword mapped  to the message that corresponds to  ``fading state''  $f(x_R)$, multiplies this symbol with $\sqrt{f(x_{R})}$, and transmits it to the destination. If ``fading state'' $f(x_R)=0 $ occurs, the source is silent and the relay disregards the received symbol. Since for $k\to\infty$,   ``fading state'' $f(x_R)$ occurs $k p^*(x_R)$ times, the source is able to transmit all of its $F$ messages in a single time slot, see  \cite{641562} for more details.

\begin{remark}\label{remark_f}
The coding performed at the EH source can be simplified  significantly    when $f(x_R)/\sigma_R^2\gg 1$  holds $\forall x_R\in \mathcal{X}_R$ for which $f(x_R)\neq 0$, using the coding scheme for the fast-fading channel in  \cite{782125, 720551}. This is because, in this case, the input distribution at the relay is uniform between $-1$ and $1$ and independent of the ``fading gain'', see (\ref{eq_e1}) and Remark~\ref{remark_3}.
As a result,  in this case, the source does not need to split $w(b)$ into $F$   messages in each time slot $b$, and thereby, does not need to use $F$ different codebooks. Instead, the source can use only one codebook  which contains $2^{k p_f R_S}$ codewords comprised of $kp_f$ symbols  where each symbol is generated independently using the uniform distribution  $p^*(v_S|x_R)$ given in (\ref{eq_pv_uniform}), and where $p_f={\rm Pr}\{f(x_R)\neq 0\}$ and $p_f R_S=R_R$ hold. Next, when the source transmits the codeword from its codebook mapped to message $w(b)$, then each  symbol of this codeword is multiplied by the ``fading state" $f(x_R)\neq 0$   corresponding to the symbol interval in which the symbol is transmitted, see \cite{782125, 720551}.  If $f(x_R)= 0$ occurs in a given symbol interval,   the source is silent and the relay disregards the received symbol.
\end{remark}

\section{Channel Capacity when the Source has an Unlimited Battery}\label{sec-4}

In the following, we determine the channel capacity of the considered relay channel for the case when the EH source is equipped with   an unlimited  battery. 

\subsection{Channel Capacity}\label{sec_4-a}

To obtain the channel capacity,  we exploit the results for the capacity of the degraded relay channel in \cite{cover} and the capacity of the EH AWGN channel with energy transmission cost in \cite{6766774}. In this way, the capacity can be obtained in a much more straightforward manner than the capacity for the case with a batteryless EH source, and can be written as
%\footnote{Note that the capacity expression in (\ref{eq_3a})   includes an average energy causality constraint  given by C1.}
\begin{align}\label{eq_3a}
C=  \max_{p(x_S,x_R)\in\mathcal{P}} \min & \left\{  I(X_S;Y_R|X_R) \; ;\; I(X_R;Y_D) \right\} \nonumber\\
\textrm{Subject to } \mathrm{C1:}& \sum_{x_S\in \mathcal{X}_S} P_S(x_S) p( x_S) \leq \sum_{x_R\in \mathcal{X}_R} E_H(x_R)  p( x_R) \nonumber\\
\mathrm{C2:} & \sum_{x_R\in \mathcal{X}_R} x_R^2 p( x_R) \leq P_R ,
\end{align} 
where $\sum_{x_R\in \mathcal{X}_R} E_H(x_R)  p( x_R) $ is the average energy harvested  by the EH source and $P_S(x_S)$ is the  energy spent   by the EH source for transmitting symbol $x_S$, which is given by  \cite{6766774}
\begin{align}\label{eq_bxr}
P_S(x_S)=\left\{
\begin{array}{ll}
0  & \textrm{if } x_S = 0\\
x_S^2+P_{\rm C} & \textrm{if } x_S\neq 0.
\end{array}
\right.
\end{align}
Hence, as can be seen from  (\ref{eq_bxr}), for every non-zero (non-silent)  symbol transmitted by the source, an additional energy $P_{\rm C}$  is needed. Constraint C1 in (\ref{eq_3a}) is due to the average energy causality constraint.
Now, since we assumed out-of-band FD relaying, which does not cause self-interference, RVs $X_{S,i}$ and $Y_{R,i}$ are both independent of RV $X_{R,i}$, cf. (\ref{eq_io_2}). As a result, the capacity expression in (\ref{eq_3a}) can be simplified as
\begin{align}\label{eq_3b}
C=   \min & \left\{ \max_{p(x_S)\in\mathcal{P}} I(X_S;Y_R)\; ;\; \max_{p(x_R)\in\mathcal{P}} I(X_R;Y_D) \right\} \nonumber\\
\textrm{Subject to } \mathrm{C1:}& \sum_{x_S\in \mathcal{X}_S} P_S(x_S) p( x_S) \leq \sum_{x_R\in \mathcal{X}_R} E_H(x_R)  p( x_R) \nonumber\\
\mathrm{C2:} & \sum_{x_R\in \mathcal{X}_R} x_R^2 p( x_R) \leq P_R .
\end{align} 
The optimal input distributions  $p^*(x_S)$ and $p^*(x_R)$  found as the solution of  (\ref{eq_3b})   depend on the  function $E_H(x_R)$. Therefore, in the following, we pursue the special case when $E_H(x_R)$ is given by (\ref{eq_E1}).

\begin{lemma}\label{lemma_3a}
In the special case when $E_H(x_R)$ is given by (\ref{eq_E1}), the capacity is given by 
\begin{align}\label{eq_3ea}
C=   \min & \left\{ I(X_S;Y_R)\Big|_{p(x_S)=p^*(x_S)} \; ; \; \frac{1}{2}\log_2\left(1+\frac{P_R}{\sigma_D^2}\right)  \right\},
\end{align} 
where $p^*(x_S)$ is discrete and found as the solution of 
\begin{align}\label{eq_3c}
 \max_{p(x_S)}\;\;& I(X_S;Y_R)  \nonumber\\
\textrm{Subject to }  
\mathrm{C1:} & \sum_{x_S\in \mathcal{X}_S} P_S(x_S) p( x_S) =  \eta h_{RS}^2 P_R,
\end{align}
where $P_S(x_S)$ is given in (\ref{eq_bxr}).
\end{lemma}
\begin{IEEEproof}
 Please refer to Appendix~\ref{app_3}.
\end{IEEEproof}

As shown in \cite{6766774}, for $P_{\rm C}>0$, the optimal input distribution at the source, $p^*(x_S)$,  found as the solution of (\ref{eq_3c}), always includes the zero (silent) symbol with non-zero probability. This means that  in order to achieve the capacity, the source is silent in a fraction of the  symbol intervals. Moreover, the source   also uses these silent symbols for encoding additional information for the relay.

\subsection{Achievability of the Channel Capacity}

The  capacity achieving coding scheme for this channel can be obtained by   combining   the coding schemes for the degraded relay channel in \cite{cover} and the coding schemes for the EH AWGN channel in \cite{6216430} and \cite{6766774}.  For completeness, we outline the combination of the coding schemes presented in \cite{6216430},   \cite{6766774}, and \cite{cover}.

The transmission is carried out in $K+N+1$ time slots, where   during each time slot  the channel is used $k$ times. The numbers  $N$ and $K$  are chosen such that    $N \to\infty$, $K\to\infty$, and $K/N\to 0$ hold. In particular, similar to the \textit{save-and-transmit} scheme in \cite{6216430}, in the first $K$ time slots, the source fills its battery without transmitting any information. To this end,  the relay sends a ``dummy'' codeword to the source in each time slot during the first  $K+1$ time slots. The destination  discards these ``dummy'' codewords received in the first $K+1$ time slots, whereas the source harvests the energy from these codewords. Next, in each time slot from the $(K+1)$-th time slot to the $(K+N)$-th time slot, the source transmit a  message  to the relay while harvesting the energy from the codeword transmitted by the relay. At the same time, the relay receives, decodes the received codewords, and then retransmits the received message in the next time slot to the destination. In the last, i.e., the $(K+N+1)$-th time slot, the source is silent and the relay retransmits to the destination the message received from the source in the previous, i.e., the $(K+N)$-th time slot. The
 transmission rates of   source and  relay in each time slot from the $(K+1)$-th to the $(N+1)$-th time slot is $C$, where $C$ is the channel capacity given in Section~\ref{sec_4-a}. The input distributions at   source and   relay are  $p^*(x_S)$ and  $p^*(x_R)$, respectively, and are  also provided in Section~\ref{sec_4-a}.   
Since $K/N\to 0$ holds, the time spent for powering up the EH source has  a negligible impact on the overall achieved data rate and on the average power consumed by the relay. Moreover,   using   the \textit{save-and-transmit} scheme in \cite{6216430}, the energy causality is satisfied for each transmitted symbol, see \cite{6216430} for a proof.

\vspace{-2mm}
\section{Numerical Examples}\label{sec-num}

In the following, we provide numerical examples to compare the  derived  capacities with several benchmark schemes. To this end, we first introduce  the system parameters,    define the  benchmark schemes, and finally provide the numerical examples.
\vspace{-2mm}
\subsection{System Parameters}\label{set_par}
We compute the channel gains of the   source-relay ($SR$) and relay-destination ($RD$) links using the standard path loss model 
\begin{align}\label{eq_h1}
h_{L}^2=\left(\frac{c }{f_{c} 4\pi}\right)^2 d_{L}^{-\alpha},\;\; \textrm{for }L\in\{SR,RD\},
\end{align}
  where $c$ is the speed of light, $f_c$ is the carrier frequency, $d_L$ is the distance between the transmitter and the receiver of link $L$, and $\alpha$ is the path loss exponent. For the numerical examples  in this section, we assume $\alpha=3$, $d_{SR}=10$ or $d_{SR}=20$  meters, and $d_{RD}=200$ meters. Moreover,   we assume  that the carrier frequencies of the transmit and receive signals of the relay are $f_c=2.3999$ GHz and $f_c=2.4001$ GHz\footnote{The  values for the carrier frequencies are chosen such that they are close to $2.4$ GHz,  which is a frequently used carrier frequency in practice. Because of the frequency separation, there is no interference between the transmit and receive  signals at the relay, which both occupy a   100 kHz bandwidth. On the other hand, $d_{RD}=200$ m is chosen to illustrate that a relatively large distance between the EH source and the destination can be bridged by using the WET transmitter as a relay.}. The  transmit bandwidth is assumed to be $B=100$ kHz. Thereby, assuming ideal Nyquist sampling, we have $2B$ independent symbols per second. Moreover, we assume that the noise power per Hz is $-160$ dBm, which leads to a total noise power of  $10^{-19} B$ Watt. 
Moreover, for the harvested energy in a single symbol interval, we assume that $E_H(x_R)$ is given in (\ref{eq_E1}) with $h_{RS}=h_{SR}$ and $\eta=0.8$. Furthermore, we assume that $P_{\rm C}=1$~mWatt. Hence,   in order for the  source to emit $P$ Watt, it has to spend an additional $1$ mWatt.
Finally, since  the  capacities derived throughout this paper are  in bits/symbol, to plot the capacities in bits/sec, we need to multiply the corresponding capacity expressions by $2B$. 

For the above  set of parameters, for the case of a batteryless EH source,  even for very small  transmit powers at the relay, $P_R$, we obtain that when  $f(x_R)$ is not zero, $f(x_R)/\sigma_R^2\gg 1$ holds.  
As a result,  we can use Corollary~\ref{cor_2} to obtain the capacity.   In particular, we obtain from  Corollary~\ref{cor_2} that the capacity is given by Case~2 in  Corollary~\ref{cor_2}, i.e., as a closed-form expression. Hence, for the adopted set of parameters, the source-relay channel is the bottleneck, which is expected in general  since the source is powered by WET, whereas the relay has its own power supply.

\vspace{-2mm}
\subsection{Benchmark Schemes}
To fairly evaluate  the capacity of the batteryless EH source and the capacity of the source with an unlimited battery,  we compare the derived capacities with  the rates achieved by several benchmark schemes as  references. For the first benchmark scheme, referred to as Benchmark Scheme~1, we assume that the source is batteryless and transmits using the optimal input distribution, given in Section~\ref{sec_pv},  however, the    relay does not use the optimal input distribution, given in Section~\ref{sec_pr},  and instead transmits   Gaussian distributed symbols, which is  a   commonly  used input distribution for the EH relay channel in the literature \cite{5958560}-\cite{7295551}. As a result, the  rate achieved by  Benchmark Scheme~1 is the minimum of the rates given by the expressions in the left hand side and the right hand side  of the inequality  in (\ref{eq_cap_2}). 
For the second benchmark  scheme, referred to as Benchmark Scheme~2, we assume
that the source is equipped with an unlimited battery. Moreover, we assume that the transmission  time is divided into slots of equal length and that one codeword spans one time slot. In addition, we assume that the source is silent for $t\geq 0$ time slots  during which it  harvests  energy from the relay and conserves it for future transmissions. Thereby, since $E_H(x_R)$ is given by (\ref{eq_E1}), in $t$ time slots  the source can harvest  $t \eta h_{RS}^2 P_R$ Watt.  Once the source has harvested enough energy, it   transmits a Gaussian distributed codeword to the relay with power $t \eta h_{SR}^2 P_R-P_{\rm C}$  spanning one time slot. Having in mind that when the source transmits information it can also harvest energy,  the maximum achievable rate on the source-relay channel using Benchmark Scheme~2 is 
\begin{align}\label{eq_0a}
\max_{t}\frac{1}{1+t}  \frac{1}{2} \log_2\left(1+\frac{(1+t) \eta h_{SR}^2 P_R-P_{\rm C}}{\sigma_R^2}\right).
\end{align}
On the other hand, for Benchmark Scheme~2, we assume that  the relay is never silent and  it transmits in each time slot a codeword  with Gaussian distributed symbols. Thereby, the maximum achievable rate on the relay-destination  channel is $(1/2) \log_2(1+P_R/\sigma_D^2 )$. Combining this rate with (\ref{eq_0a}), we obtain the  maximum achievable rate for Benchmark Scheme~2 as
\begin{align}\label{eq_0}
R=\min&\left\{\max_{t} \frac{1}{1+t} \frac{1}{2} \log_2\left(1+\frac{(1+t) \eta h_{SR}^2 P_R-P_{\rm C}}{\sigma_R^2}\right)\;,\; \right. \; \left. \frac{1}{2} \log_2\left(1+\frac{P_R}{\sigma_D^2}  \right)\right\}.
\end{align}
Finally,   we also use the rate achieved with the protocol in \cite{7088643}  as   a benchmark. However, for   fair comparison, we modify the  scheme proposed in  \cite{7088643} and instead of a half-duplex relay   we assume an out-of-band FD relay. According to the protocol in \cite{7088643}, both source and relay transmit Gaussian distributed symbols. Thereby, the achievable rate of this benchmark scheme, denoted by Benchmark Scheme~3, is identical to    (\ref{eq_0}) with the parameter $t$ set to zero. Hence, Benchmark Scheme~3 is identical to Benchmark Scheme~2 if the source is not allowed to conserve energy but is forced to transmit   Gaussian signals in each time slot using.

\subsection{Numerical Results}

\begin{figure}
\centering
\includegraphics[width=5in]{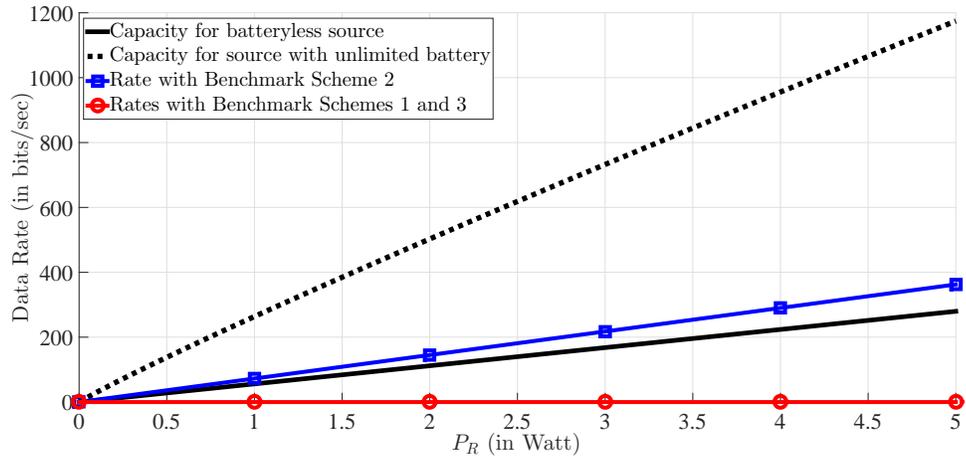}
 %\vspace*{-3mm}
\caption{Comparison of capacities and achievable rates of the benchmark schemes  as a function of the relay's power $P_R$ in Watt  for $d_{SR}=10$ meter.}  \label{fig_3}
 \vspace*{-3mm}
\end{figure}

For the above set of parameters with a source-relay distance of $d_{SR}=10$ meter, the channel capacities of the considered relay channel with a batteryless EH source and an EH source  with an unlimited battery, respectively, are shown in Fig.~\ref{fig_3}. In addition,  in Fig.~\ref{fig_3}, we also show  the rates achieved with the three benchmark schemes. As can be seen from Fig.~\ref{fig_3}, for $P_R$ in the range from zero to five Watts,  the channel capacity for the case when the source is equipped with an unlimited battery is more than four times higher  than the channel capacity for the case when the source is batteryless. This is expected since, in the former case, the source can store energy in its battery and then use it later   to transmit information to the relay. In fact, for the capacity of the EH source with unlimited battery,  in a large portion of the transmission time, the EH source is silent and conserves energy. Moreover, the EH source   uses the  silent symbols  for encoding additional information for the relay. Although the batteryless EH source is also silent in a large portion of the symbol intervals, it cannot use the    silent symbols  for encoding additional information for the relay since the relay knows when these silent symbol intervals occur.   On the other hand, in   Benchmark Scheme~2, the source is also silent during $t$ time slots and   conserves energy. However, since for Benchmark Scheme~2 the source does not use the silent symbols for encoding additional information for the relay, the rate of Benchmark Scheme~2, given by (\ref{eq_0}),  is much lower than the derived   capacity when the source has an unlimited battery  and is only slightly larger than the derived capacity when the source is batteryless. This shows that the encoding of information in the silent symbols at the EH source with a battery has a large impact on the achievable data rate. 
Moreover, Fig.~\ref{fig_3}  also shows that using the optimal coding scheme with the optimal input  distributions at the source and the relay is essential for high performance for both battery scenarios.
For example, if  non-optimal Gaussian signaling is used at the relay for the case of the batteryless EH source,  as in Benchmark Scheme~1, or if   non-optimal Gaussian signaling is used at the source   for the case when the source is equipped with an unlimited battery, as in Benchmark Scheme~3,  the  data rate  is zero for the adopted range of $P_R$, i.e., no information can be transmitted by the source to the relay for Benchmark Schemes 1 and 3 for $P_R\leq 5$ Watt. The poor performance is a consequence of the fact that Benchmark Schemes 1 and 3 do not take into  account the  energy transmission cost. More precisely, for Benchmark  Scheme~1, due to the non-optimal signaling used by   the relay, the source is not able to  harvest more energy than what is consumed by the energy transmission cost $P_{\rm C}$. Hence, since every attempt to emit a non-zero symbol incurs an energy transmission cost, the energy left for information transmission  after subtracting the energy transmission cost  is zero.
 On the other hand, for Benchmark  Scheme~3, due to the non-optimal signaling used by the source, the source is forced to transmit a non-zero symbol in every symbol interval  without   having the chance to be silent and to conserve energy. Due to the energy transmission   cost, the energy left for information transmission     is again zero for $P_R\leq 5$ Watt.

\begin{figure}
\centering
\includegraphics[width=5in]{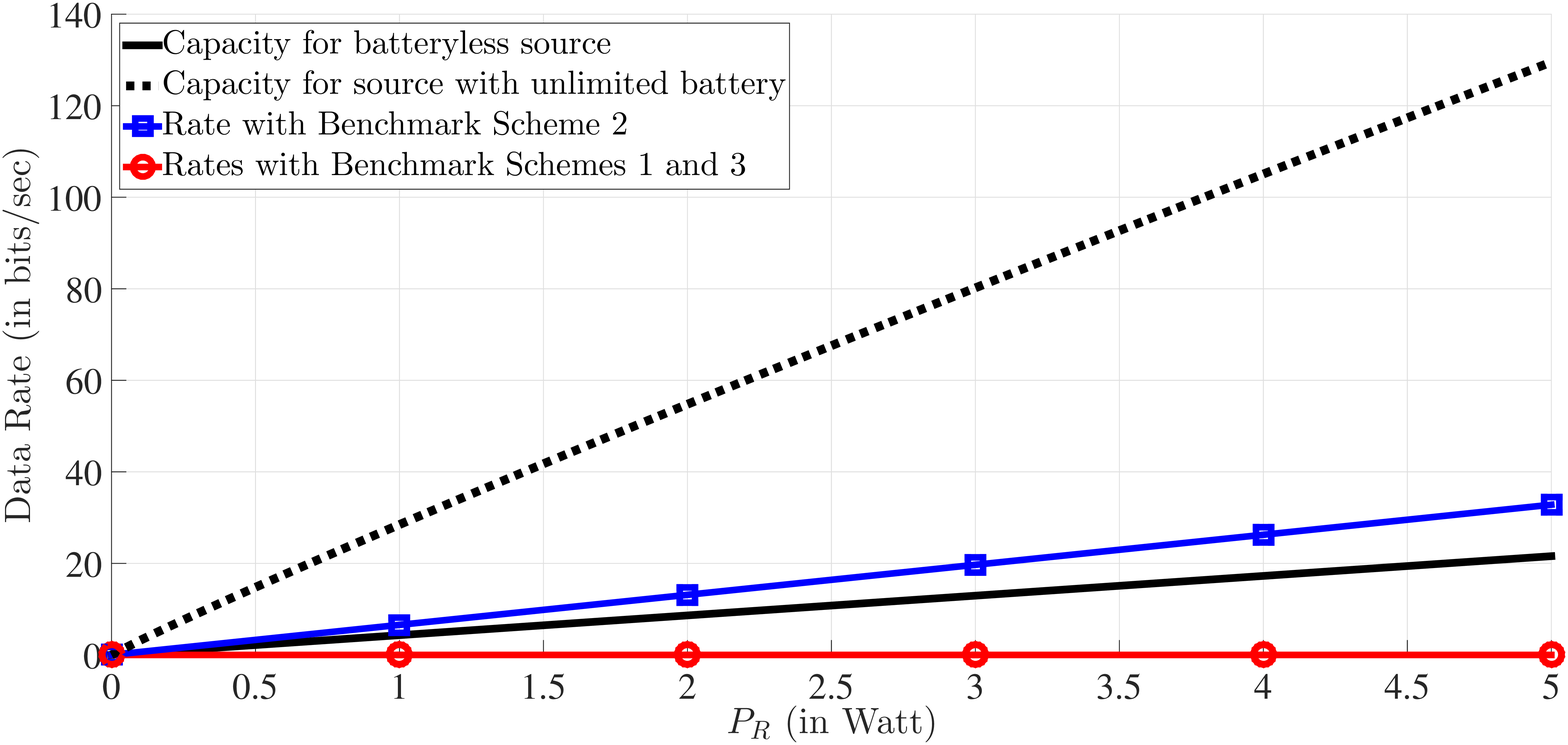}
 %\vspace*{-3mm}
\caption{Comparison of capacities and achievable rates of the benchmark schemes  as a function of the relay's power $P_R$ in Watt  for $d_{SR}=20$ meter.}  \label{fig_3a}
 \vspace*{-3mm}
\end{figure}

For Fig.~\ref{fig_3a}, we use the same parameters as in  Fig.~\ref{fig_3} but  the  distance  between the source and relay is increased to $d_{SR}=20$ meter.  Comparing Fig.~\ref{fig_3} and  Fig.~\ref{fig_3a}, we can see that the doubling of the source-relay distance results in a tenfold  reduction of the channel capacity for both the batteryless EH source and the source with an unlimited battery.

 \begin{figure}
\centering
\includegraphics[width=5in]{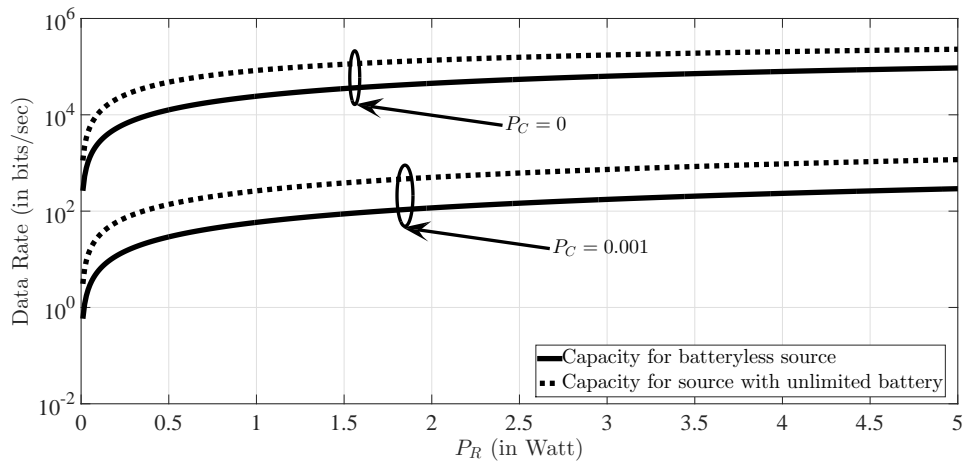}
 %\vspace*{-3mm}
\caption{Comparison of capacities for different energy transmission costs $P_{\rm C}$  as a function of the relay's power $P_R$ in Watt.}  \label{fig_4}
 \vspace*{-3mm}
\end{figure}

To illustrate the effect that the energy transmission cost has on the channel capacities,   in Fig.~\ref{fig_4}, we show the capacities\footnote{We note that the rates in Figs.~\ref{fig_3} and \ref{fig_3a}  appear to be a linear  function of $P_R$ since the source-relay channel operates in the low  SNR regime due to the high-path loss attenuation associated with WET. The rates in Fig.~\ref{fig_4} are also a linear   function of $P_R$, however, since the y-axis in Fig.~\ref{fig_4}  is given in the logarithmic scale, this linearity is not obvious.} for $P_{\rm C}=1$ mWatt  and  $P_{\rm C}=0$ Watt (zero energy transmission cost\footnote{For the adopted set of parameters, in the case when $P_{\rm C}=0$, the expression for the channel capacity  for the batteryless case  is given by Case~1 in Corollary~\ref{cor_2} with  $P_{\rm C}=0$, whereas, the capacity for the case of a source with unlimited battery is given by (\ref{eq_3ea}) with $I(X_S;Y_R)\big|_{p(x_S)=p^*(x_S)} = \frac{1}{2}\log_2\left(1+\frac{\eta h_{SR}^2 P_R}{\sigma_D^2}\right)$.}), for the case when the distance between the source and the relay is $d_{SR}=10$  meter. The figure shows that  a non-zero energy cost has a  severe impact on the channel capacity. In particular, for the considered parameters, the capacities for $P_{\rm C}=0$ Watt are approximately $10^3$ times higher than the capacities for $P_{\rm C}=1$ mWatt.
 Hence,  for the considered relay channel, any approximation of the achievable data rates made by  neglecting the energy transmission cost can result in a severe overestimation of the achievable performance.

\section{Conclusion}\label{sec-conc}

We have derived the capacity of a  two-hop relay channel impaired by AWGN, where an EH source   is powered wirelessly by an out-of-band FD relay. 
  We assumed that the relay has an average transmit power constraint whereas the source has an  energy transmission cost constraint. Moreover, we considered  two extreme cases for the battery at the EH source,  a batteryless  source and a source equipped with an unlimited battery. For both cases, we showed that  in order   to achieve the capacity of the considered relay channel,   the source has to harvest the RF energy that reaches the source when the relay transmits information to the destination.   Moreover, for both considered cases, we demonstrated that the   capacity-achieving distribution at the source is discrete, whereas the capacity-achieving distribution at the relay can either  be discrete or zero-mean Gaussian.  Furthermore, our results revealed  that the use of suboptimal input distributions may incur a severe  degradation in   performance, and neglecting the energy transmission cost at the source can result in a severe overestimation of the achievable performance.

\appendix

\subsection{Proof of Lemma~\ref{lemma_1a}}\label{app_1a}
Since the considered relay channel belongs to the class of degraded relay channels, its capacity is given  by the capacity expression for the degraded relay channel in \cite{cover}. Taking  the amplitude constraint at the source, given by  (\ref{eq_2b}), and the average power constraint at the relay,  $P_R$, into account, the capacity   can be expressed in the following general form
\begin{align}\label{eq_3}
C=  \max_{p(v_S,x_R)\in \mathcal{P}} \min \{ &I(V_S;Y_R|X_R,f(X_R));I(X_R;Y_D)\} \nonumber\\
\textrm{Subject to } \mathrm{C1:}& -1 \leq  V_S   \leq 1, \nonumber \\
\mathrm{C2:} & \sum_{x_R\in \mathcal{X}_R} x_R^2 p( x_R) \leq P_R .
\end{align}
Now, since $f(X_R)$ is a deterministic function of $X_R$, conditioning the mutual information on $X_R$ and $f(X_R)$ is equivalent to conditioning only on $X_R$. As a result, the conditioning on $f(X_R)$ in (\ref{eq_3}) can be removed.
To further simplify the capacity expression in (\ref{eq_3}),   note that $p(v_S,x_R)$ can be written as $p(v_S,x_R)=p(v_S|x_R)p(x_R)$. As a result,   the maximization over $p(v_S,x_R)$ can be replaced by two nested maximizations, one with respect to $p(v_S|x_R)$ for a fixed $p(x_R)$, and the other one with respect to $p(x_R)$. Thereby, (\ref{eq_3}), with the conditioning on $f(X_R)$ removed, can be written equivalently as
\begin{align}\label{eq_4a}
C=  \max_{p(x_R)\in \mathcal{P}} \max_{p(v_S|x_R)\in \mathcal{P}} \min \{&I(V_S;Y_R|X_R);I(X_R;Y_D)\} \nonumber\\
\textrm{Subject to } \mathrm{C1:}& -1 \leq  V_S   \leq 1,  \nonumber\\
\mathrm{C2:} & \sum_{x_R\in \mathcal{X}_R} x_R^2 p( x_R) \leq P_R .
\end{align}
Now,  in the capacity expression in (\ref{eq_4a}), note that  only $I(V_S;Y_R|X_R)$ 
depends on   $p(v_S|x_R)$ whereas $I(X_R;Y_D)$ is not dependent on $p(v_S|x_R)$. As a result,  the capacity expression in   (\ref{eq_4a})  can be further  simplified as in (\ref{eq_5}),  where we have exploited the   identity
$
I(V_S;Y_R|X_R)=\sum\limits_{x_R\in \mathcal{X}_R}   I(V_S;Y_R|X_R=x_R) p(x_R).
$

\subsection{Proof of Lemma~\ref{lemma_2a}}\label{app_1b}
The optimization problem in (\ref{eq_5}) with respect to $p(v_S| x_R)$  can be resolved into the following, much simpler, optimization problem
\begin{align}\label{eq_5as}
 \max_{p(v_S|x_R)\in \mathcal{P}} &\; I(V_S;Y_R|X_R=x_R)  \nonumber\\
\textrm{Subject to }  \mathrm{C1: }  &\;-1 \leq  V_S   \leq 1. 
\end{align}
On the other hand, from \cite{smith1971information}  it is known that the optimal input distribution  that maximizes the mutual information of a point-to-point AWGN channel  with a fixed channel gain and an amplitude constraint imposed at the transmitter  is discrete with a finite number of probability mass points. Since $I(V_S;Y_R|X_R=x_R)$ in (\ref{eq_5as}) is the  mutual information of the source-relay AWGN channel for a fixed channel gain $\sqrt{f(X_R=x_R)}$  and an  amplitude constraint given by  C1 in (\ref{eq_5as}),  the optimal input distribution $p^*(v_S|x_R)$ obtained as the solution of (\ref{eq_5as})  is discrete with a finite number of probability mass points. As a result, $p^*(v_S|x_R)$ can be written in a general form as (\ref{eq_pv}).

Now,  to obtain $\max\limits_{p(v_S|x_R)\in \mathcal{P}} I(V_S;Y_R|X_R=x_R)$, which is $ I(V_S;Y_R|X_R=x_R)$ for $p^*(v_S|x_R)$ given in (\ref{eq_pv}), we use the following identity
\begin{align}\label{eq_eq_ce}
 I(V_S;Y_R|X_R=x_R) = h(Y_R|X_R=x_R)-h(Y_R|V_S,X_R=x_R),
\end{align}
where $h(\cdot|\cdot)$ denotes the conditional differential entropy \cite{cover2012elements}. In (\ref{eq_eq_ce}),  $h(Y_R|V_S,X_R=x_R)$ is  the differential entropy of the AWGN at the relay, cf.~Fig.~\ref{fig_2}, which is given by
\begin{align}\label{eq_eq_cea}
h(Y_R|V_S,X_R=x_R)=\frac{1}{2} \log_2\big(2\pi e \sigma_R^2\big).
\end{align}
On the other hand, we can obtain $h(Y_R|X_R=x_R)$ by definition as \cite{cover2012elements} 
\begin{align}\label{eq_saed}
h(Y_R|X_R=x_R) = -\int_{-\infty}^\infty \sum_{v_{S}\in V_{S} } p(y_R|v_S,x_R) p(v_S|x_R) \log_2\left(\sum_{v_{S}\in V_{S}}  p(y_R|v_S,x_R) p(v_S|x_R)  \right ) d y_R,
\end{align}  
where the optimal $p(v_S|x_R)$ is given by (\ref{eq_pv}) and  $p(y_R|v_S,x_R)$ is a  Gaussian distribution with variance $\sigma_R^2$ and mean $v_S \sqrt{f(x_R)}$, cf.~Fig.~\ref{fig_2}. Inserting   (\ref{eq_eq_cea})  and (\ref{eq_saed}) into (\ref{eq_eq_ce}), we   finally obtain  $ \max\limits_{p(v_S|x_R)\in \mathcal{P}}I(V_S;Y_R|X_R=x_R)$  as in (\ref{eq_Is}).

\subsection{Proof of Theorem~\ref{theo_1}}\label{app_1}

The maximization problem in (\ref{eq_5})  can be written equivalently using the epigraph form as 
\begin{align}\label{MPR1}
\begin{array}{rl}
  {\underset{u,\; p(x_R)}{\rm{Maximize}}} \;&  u \\
 {\rm{Subject\;\; to \;\; }} 
   {\rm C1:}\;&  u- \sum\limits_{x_R\in \mathcal{X}_R} \max\limits_{p(v_S|x_R)\in \mathcal{P} } I(V_S;Y_R|X_R=x_R) p(x_R)  \leq 0   \\
   {\rm C2:}\; &  u- I(X_R;Y_D)\leq 0  \\
  {\rm C3:}\; &  \sum\limits_{x_R\in \mathcal{X}_R}  x_R^2 p(x_R)  -P_R \leq 0\\
 {\rm C4:}\;  & \sum\limits_{x_R\in \mathcal{X}_R}    p(x_R)  - 1 =  0.
\end{array} 
\end{align} 

The optimization problem in (\ref{MPR1}) is a concave optimization problem since constraints C1, C3, and C4 are all affine with respect to $p(x_R)$ and constraint C2 is convex with respect to $p(x_R)$. Hence,   (\ref{MPR1}) can be solved using the Lagrangian method \cite{Boyd_CO}. The Lagrange function for optimization problem (\ref{MPR1}) is given by
\begin{align}\label{eq_ap2}
\mathcal{L}  &= u-\alpha_1 \left( u-  \sum\limits_{x_R\in \mathcal{X}_R}   \max\limits_{p(v_S|x_R)\in\mathcal{P} } I(V_S;Y_R|X_R=x_R) p(x_R)   \right)   -\alpha_2 \left( u- I(X_R;Y_D) \right)\nonumber\\ & -\lambda \left( \sum\limits_{x_R\in \mathcal{X}_R} \hspace{-1mm} x_R^2 p(x_R)   -P_R\right) -\xi \left(     \sum\limits_{x_R\in \mathcal{X}_R}\hspace{-1mm} p(x_R)  -1\right ) ,
\end{align} 
where $\alpha_1$, $\alpha_2$, $\lambda$, and $\xi$ are Lagrange multipliers, which have to satisfy the following Karush-Kuhn-Tucker (KKT) conditions 
\begin{subequations}
\begin{align}
&\alpha_1 \left( u- \sum\limits_{x_R\in\mathcal{X}_R} \max\limits_{p(v_S|x_R)\in \mathcal{P} } I(V_S;Y_R|X_R=x_R) p(x_R) \right)=0  \textrm{ and }\alpha_1\geq 0 ,\label{eq_ap4} \\
&\alpha_2 \left( u- I(X_R;Y_D) \right)=0 \textrm{ and }\alpha_2\geq 0,  \label{eq_ap5} \\
&\lambda \left(\sum\limits_{x_R\in \mathcal{X}_R}  x_R^2 p(x_R)-P_R\right ) =0 \textrm{ and }\lambda\geq 0,  \quad
 \xi \left( \sum\limits_{x_R\in \mathcal{X}_R}  p(x_R)-1\right )=0   \textrm{ and } \xi \geq  0   .\label{eq_ap6a}
\end{align}
\end{subequations}
Differentiating $\mathcal{L} $ with respect to $u$ and equating the result to zero, we obtain that $\alpha_1=1-\alpha_2=\alpha$ has to hold in order to have  a bounded solution for the dual problem (\ref{eq_ap2}), where $0\leq\alpha\leq 1$. To obtain the maximum of $\mathcal{L} $ with respect to $p(x_R)$,  we need to obtain the   derivative of $\mathcal{L}$ with respect to $p(x_R)$, denoted by $\partial \mathcal{L} /\partial p(x_R)$, and  equate it to zero. Thereby, we obtain 
\begin{align}\label{eq_ap7}
 & \alpha  \max\limits_{p(v_S|x_R)\in\mathcal{P}} I(V_S;Y_R|X_R=x_R)   - (1-\alpha)  I'(X_R;Y_D)    -\lambda   x_R^2 -\xi =0,
\end{align} 
where $ I'(X_R;Y_D)=\frac{\partial}{\partial p(x_R)} I (X_R;Y_D) $  is given by
\begin{align}\label{eq_dfg}
I'(X_R;Y_D) =\int_{y_D} p(y_D|x_R)p(x_R)\log_2\left(\frac{p(y_D|X_R)}{p(y_D)}\right)-\frac{1}{\ln(2)}.
\end{align} 
  We note that there are  three possible solutions for (\ref{eq_ap7}) depending on whether $\alpha= 0$,  $\alpha=1$, or $0<\alpha< 1$, respectively. In the following, we analyze these solutions.

 If $\alpha=0$,  we obtain  from  (\ref{eq_ap4}) that for the optimal $p^*(x_R)$,   the following has to hold
\begin{align}\label{eq_ap8}
u   <  \sum\limits_{x_R\in \mathcal{X}_R} \max\limits_{p(v_S|x_R)\in\mathcal{P}} I(V_S;Y_R|X_R=x_R) p^*(x_R) \quad \textrm{ and }  \quad u =I(X_R;Y_D)\Big|_{ p(x_R)= p^*(x_R)} , 
\end{align}
which is equivalent to
\begin{align}\label{eq_ap10}
u&= I(X_R;Y_D)\Big|_{ p(x_R)= p^*(x_R)}   < \sum\limits_{x_R\in \mathcal{X}_R} \max\limits_{p(v_S|x_R)\in\mathcal{P}} I(V_S;Y_R|X_R=x_R) p^*(x_R).
\end{align}
Hence, when $\alpha=0$, in order to maximize $u$, we need to maximize $I(X_R;Y_D)$. As a result,  the optimal $p^*(x_R)$ in this case is found as the distribution which maximizes $I(X_R;Y_D)$ with the average power constraint $P_R$ imposed. For the AWGN channel, this distribution is   known and is the zero-mean Gaussian distribution with variance $P_R$, for which $I(X_R;Y_D)$ is given by \cite{cover2012elements}
\begin{align}\label{eq_asa}
I(X_R;Y_D)=\frac{1}{2}\log_2\left(1+\frac{P_R}{\sigma_D^2}\right).
\end{align}
Averaging $I(V_S;Y_R|X_R=x_R)$, given in (\ref{eq_Is}), with respect to   the zero-mean Gaussian distribution with variance $P_R$, and inserting the result along with (\ref{eq_asa}) into (\ref{eq_ap10}),  we obtain Case 2 in Theorem~\ref{theo_1}.

On the other hand, if $\alpha=1$,  we obtain from  (\ref{eq_ap4})  that  for the optimal $p(x_R)$, the following has to hold
\begin{align}\label{eq_ap13}
u   = \sum\limits_{x_R\in \mathcal{X}_R} \max\limits_{p(v_S|x_R)\in\mathcal{P}} I(V_S;Y_R|X_R=x_R) p^*(x_R)    \quad \textrm{ and }  \quad
u   <I(X_R;Y_D) \Big|_{ p(x_R)= p^*(x_R)} ,
\end{align}
 which is equivalent to
\begin{align}\label{eq_ap15}
u&= \sum\limits_{x_R\in \mathcal{X}_R} \max\limits_{p(v_S|x_R)\in\mathcal{P}} I(V_S;Y_R|X_R=x_R) p^*(x_R)  < I(X_R;Y_D)\Big|_{ p(x_R)= p^*(x_R)}.
\end{align}
Hence, when $\alpha=1$, in order to maximize $u$, we need to maximize $\sum\limits_{x_R\in \mathcal{X}_R} \max\limits_{p(v_S|x_R)} I(V_S;Y_R|X_R=x_R) p(x_R)$ and for the resulting $p^*(x_R)$,  (\ref{eq_ap15}) should hold, which leads to Case~1 in Theorem~\ref{theo_1}.

Finally, if $0<\alpha<1$, then   for the optimal $p^*(x_R)$ the following holds
\begin{align}\label{eq_ap17}
u &= I(X_R;Y_D)\Big|_{ p(x_R)= p^*(x_R)}  =  \sum\limits_{x_R\in \mathcal{X}_R} \max\limits_{p(v_S|x_R)\in\mathcal{P}} I(V_S;Y_R|X_R=x_R) p^*(x_R).
\end{align} 
Considering  that $\alpha$ in (\ref{eq_ap7}) satisfies $0<\alpha<1$, we can  write (\ref{eq_ap7}) equivalently  as
\begin{align}\label{eq_ap18}
   I'(X_R;Y_D)  &= -\frac{\alpha}{1-\alpha}  \max\limits_{p(v_S|x_R)\in\mathcal{P}} I(V_S;Y_R|X_R=x_R)  +    \frac{\lambda}{1-\alpha}   x_R^2 + \frac{\xi}{1-\alpha}.
\end{align} 
Now, using the approach in \cite{1459046} it can be shown that the $p^*(x_R)$ that satisfies (\ref{eq_ap18}) cannot be a continuous distribution and can only  be discrete. Combining this with (\ref{eq_ap17}), we obtain Case~3 in Theorem~\ref{theo_1}.

\begin{remark}\label{rem_9}
Although  we derived  (\ref{eq_ap7}) assuming that $p(x_R)$ is discrete, we would have  arrived at the same result if we had assumed that $p(x_R)$ was a continuous distribution. To this end, we first would have to replace the sums in the optimization problem in (\ref{MPR1}) with integrals with respect to $x_R$. Next, in order to obtain the stationary points of the  corresponding Lagrangian function, instead of the ordinary derivative, we would have to take the functional derivative and equate it to zero. This again would lead to the identity in (\ref{eq_ap7}). Hence, the conclusions drawn from  the Lagrangian and   (\ref{eq_ap7}) are also valid when  $p(x_R)$ is a  continuous distribution. 
\end{remark}

\subsection{Proof of Corollary~\ref{cor_2}}\label{app_2}
Corollary~\ref{cor_2} follows by solving (\ref{eq_cap_1}) for $ \max\limits_{p(v_S|x_R)\in\mathcal{P}}I(V_S;Y_R|X_R=x_R) $ given in (\ref{eq_eq1}) and $E_H(x_R)$ given in (\ref{eq_E1}). The corresponding Lagrangian of this optimization problem is 
\begin{align}\label{eq_h}
 \mathcal{L}&= \sum\limits_{x_R\in \mathcal{X}_R} \frac{1}{2}\log_2\left( 1+\frac{2 \max\big(0,\eta  h_{RS}^2 x_{R}^2- P_{\rm C} \big) }{\pi e \sigma_R^2} \right) p(x_R) -\lambda\left(  \sum_{x_R\in \mathcal{X}_R} x_R^2 p( x_R) - P_R\right)\nonumber\\
& -\xi \left(\sum_{x_R\in \mathcal{X}_R}  p( x_R) -1\right).
\end{align}
Differentiating $ \mathcal{L}$ with respect to $p( x_R)$ and equating the result to zero, we obtain that for $0< p(x_R)<1$, the following has to hold
\begin{align}\label{eq_sde6}
  \frac{1}{2}\log_2\left( 1+\frac{2 \max\big(0,\eta  h_{RS}^2 x_{R}^2- P_{\rm C} \big)}{\pi e \sigma_R^2} \right)=\lambda     x_R^2   +\xi  .
\end{align}
Now, from (\ref{eq_sde6}) we can see that if $\xi\neq 0$,  then (\ref{eq_sde6}) does not hold for $x_R=0$. Furthermore, for    $\xi\neq 0$,  (\ref{eq_sde6}) has only two solutions which are in the form $\pm x^*_{R}$. Since there are only two solutions for $x_R$, in order for $p(x_R)$ to be a valid distribution, $p(x^*_{R})=1-p(-x^*_{R})$ has to hold. Moreover, in order for C1 in (\ref{eq_cap_1}) to hold  $x^*_{R}=\sqrt{P_R}$ has to hold. Hence, one possible solution for $p(x_R)$ is given in  (\ref{eq_dfs1}) and this solution is possible if 
\begin{align}\label{eq_sde7}
  \frac{1}{2}\log_2\left( 1+\frac{2 \max\big(0,\eta  h_{RS}^2 P_{R}- P_{\rm C} \big)}{\pi e \sigma_R^2} \right)=\lambda     P_R   +\xi  
\end{align}
holds or equivalently if  $\eta  h_{RS}^2 P_{R}> P_{\rm C} $ holds.   
On the other hand, if  $\eta  h_{RS}^2 P_{R}> P_{\rm C}$ does not hold, then $\xi=0$ and the following has to hold
\begin{align}\label{eq_sde8}
  \frac{1}{2}\log_2\left( 1+\frac{2 \max\big(0,\eta  h_{RS}^2 x_{R}^2- P_{\rm C} \big)}{\pi e \sigma_R^2} \right)=\lambda     x_R^2    .
\end{align}
Now,   (\ref{eq_sde8}) can be solved in closed form. In particular,  we obtain three  solutions  for $x_R$, $x_R=0$,  $x_R=-x_0$, and  $x_R=x_0$, where $x_0$ is given in (\ref{eq_sde9a}). For these three values of $x_R$, we can find   $p(x_R)$   from constraint  C1 in (\ref{eq_cap_1}) and from the constraint that $p(x_R)$  has to be a valid probability distribution. This concludes the proof.

\subsection{Proof of Lemma~\ref{lemma_3a}}\label{app_3}

If  $E_H(x_R)$ is given by (\ref{eq_E1}), then constraint C2 in (\ref{eq_3b}) has to hold with equality since $\max\limits_{p(x_R)}I(X_R;Y_D) $ is a non-decreasing function of $P_R$. As a result, we obtain the right hand side of   constraint C1 in (\ref{eq_3b}) as  $\sum_{x_R\in \mathcal{X}_R} E_H(x_R)  p( x_R)=\eta h_{RS}^2 P_R$. Since $\max_{p(x_S)} I(X_S;Y_R)$ is also a non-decreasing function of $P_R$, constraint C1 in (\ref{eq_3b}) also has to hold with equality. Consequently, the optimization problem in (\ref{eq_3b}) can be decomposed into  two optimization problems. The first optimization problem is $ \max_{p(x_R)}   I(X_R;Y_D)$ subject to  $\sum_{x_R\in \mathcal{X}_R} x_R^2 p( x_R) = P_R$, whose   solution for  $p^*(x_R)$ is   the zero-mean Gaussian distribution with variance $P_R$ and consequently $\max\limits_{p(x_R)\in\mathcal{P}} I(X_R;Y_D) =1/2\log_2(1+P_R/\sigma_D^2)$.  
On the other hand, the second optimization problem is given by  (\ref{eq_3c}). It is proven in \cite{6766774} that the optimal distribution $p^*(x_S)$,  obtained as the solution of (\ref{eq_3c}),   is discrete. As a result, this distribution can   be found by solving the concave optimization problem in (\ref{eq_3c}) numerically using   numerical optimization software such as Mathematica.   Combining the above results, for    $E_H(x_R)$   given  by (\ref{eq_E1}), we obtain the capacity as (\ref{eq_3ea}).

%%%%%%%%%%%%%%%%%%%%%%%%%%%%%%%%%%%%%%%%%%%
\bibliography{litdab}
\bibliographystyle{IEEEtran}
% that's all folks

\end{document}